\documentstyle[12pt,epsfig]{article}
\newcommand{\beq}{\begin{equation}}
\newcommand{\eeq}{\end{equation}}

\def\fun#1#2{\lower3.6pt\vbox{\baselineskip0pt\lineskip.9pt
\ialign{$\mathsurround=0pt#1\hfil##\hfil$\crcr#2\crcr\sim\crcr}}}

\begin{document}
{\title{Low- and Intermediate-Energy Nucleon-Nucleon Interactions and
the Analysis of Deuteron Photodisintegration within the Dispersion
Relation Technique}
\vskip 1cm
\author{ A.V.Anisovich and
V.A.Sadovnikova}
\date{ }
\maketitle

\vskip 1cm

\begin{abstract}
The nucleon-nucleon interaction in the region of the nucleon
kinetic energy up to $1000$ MeV is analysed together with the
reaction $\gamma d \to pn$ in the photon energy range
$E_{\gamma}=0-400$ MeV.  Nine  nucleon-nucleon $s$-channel partial
amplitudes are reconstructed in the dispersion relation
$N/D$ method: $^1S_0$,
$^3S_1-^3D_1$, $^3P_0$, $^1P_1$, $^3P_1$, $^3P_2$, $^1D_2$, $^3D_2$
and $^3F_3$. Correspondingly, the dispersive representation of partial
amplitudes  $N\Delta \to pn$, $NN^* \to pn$ and $NN\pi \to pn$ is
given. Basing on that, we have performed parameter-free calculation of
the amplitude $\gamma d \to
pn$,  taking into account: $(i)$ pole diagram,  $(ii)$
nucleon-nucleon final-state rescattering $\gamma d \to pn \to pn$, and
$(iii)$ inelastic final-state rescatterings $\gamma d  \to
 N\Delta(1232) \to pn$, $\gamma d  \to NN^*(1400) \to pn$ and $\gamma d
\to NN\pi \to pn$. The
  $\gamma d \to  pn$ partial amplitudes for
 nine above-mentioned channels are found.  It is shown that the process
$\gamma d \to pn  \to pn$ is significant for the waves  $^1S_0$,
$^3P_0$, $^3P_1$, at $E_{\gamma} =50 -100$ MeV, while $\gamma d \to
 N\Delta \to pn$ for the waves  $^3P_2$, $^1D_2$,$^3F_3$ dominates at
$E_{\gamma} > 300$ MeV.  Meson exchange current contributions into the
deuteron disintegration are estimated:  they are significant at
$E_\gamma =100-400$ MeV.

\end{abstract}

\bigskip

\section{Introduction}

Investigations of the lightest nuclei such as deuteron, $^3He$,
$^3H$, or $^4He$ were always challenging for researchers.
There are  at least  two reasons  for that:\\
$(i)$ Investigations of simple nuclei give a chance to comprehend
better the crucial problems of nuclear physics, such as the
 structure of nucleon-nucleon forces,  the role  of multinucleon
interactions, the role of internuclear forces at small and large
distances, the problem of multiquark bags like $6q$ or $12q$, meson
degrees of freedom in nuclei, the role of relativistic effects, etc.
\\$(ii)$ The study of light nuclei
provides an opportunity to elaborate the methods for the description
of composite systems.  Hadron physics is a physics of composite
particles --- quarks and gluons being constituents.
Elaboration of new approaches for the description of composite systems
based on the study of light nuclei has an undoubtful advantage that
constituents (nucleons) are observed particles, with known
interaction amplitudes.

This paper is devoted to simultaneous investigation of the two
two-nucleon processes, namely, $NN \to NN$
at $0 \le T \le 1000$ MeV and $\gamma d \to pn$ at
$0 \le E_{\gamma} \le 400$ MeV. This simultaneous investigation is not
accidental: the $NN \to NN$ amplitude is basic for the study
of the two-nucleon reactions. At the same time, the reaction  $\gamma d
\to pn$ is one of the simplest processes with two nucleons; therefore,
the simultaneous analysis should clarify the points:\\
$(i)$ To what extent and at what energies the nucleon-nucleon
interaction governs the other processes induced by this interaction;
and
\\ $(ii)$ What characteristics of the nucleon-nucleon interaction
play the leading role.

For the deuteron
photodisintegration at the photon energies  in the laboratory
frame  $0 \le E_{\gamma} \le 400$ MeV,   numerous
    experimental data exist, see refs.
\cite{mor} -- \cite{meyer} and references therein. This
region covers both small energies, where
quantum mechanics works, and
comparatively large energies, where
relativistic effects and inelastic processes are important.

The reaction $\gamma d \to pn$ is investigated using
dispersive integration technique, which is appropriate for the
analysis of partial wave amplitudes:  \\
$(i)$ This
technique is  relativistic invariant;\\
$(ii)$ It keeps under control all
 intermediate states allowing to take properly into account final
state rescatterings which are determined by
 the right-hand singularities of the  amplitudes \cite{chewmand};
$(iii)$ In this technique there is no ambiguity related to the
mass-off-shell amplitudes.

The dispersion relation technique has much common with both
non-relativistic quantum mechanics approach and light-cone variable
technique, the methods which are widely used by non-perturbative QCD
(or Strong-QCD \footnote{This terminology is suggested by F.E.
Close \cite{fec}}).  A similarity of the dispersion
relation and light-cone
variable techniques was specially  outlined in ref.  \cite{amn}
where  the Perturbative-QCD
onset in meson form factors has been analysed.

Consider in more detail the mechanism
of the deuteron photodisintegration,
 $\gamma d \to pn$, at $E_{\gamma} \le 400$ MeV.
The following processes contribute to this
reaction: Fig. 1a ---  the deuteron break-up without
nucleon interactions in the final state, Fig. 1b --- the deuteron break-up
followed by the rescattering of outgoing nucleons,
Fig. 1c,d --- the photoproduction of one or several mesons in the
intermediate state, or --- Fig. 1e,f --- the photoexcitation of
resonances, like $\Delta$ and $N^*(1400)$,  followed by the transition
of the produced particles into  the $pn$ state.

We start with the discussion of the  nucleon--nucleon interaction.
There is a striking property relevant to the hadron interaction at low
and intermediate energies: in a wide energy range
the nucleon interaction is almost elastic or quasi-elastic
(this means an excitation of nucleons into highly located baryon
states), i.e. with the energy growth the onset of
 genuine inelastic processes is delayed. Typical example is
given by the deuteron channel ($pn$ scattering in the $^3S_1$ and
$^3D_1$ waves) where, though the threshold of the pion production is at
$T=290$ MeV,
 the phase shift analysis indicates that the inelasticity
parameter is strongly suppressed up to $T=1000$ MeV \cite{arndt}. For
other waves the inelasticities are related to baryon excitations; for
instance, in the $^1D_2$ and $^3F_3$ waves the inelasticities are due
to the transitions $NN \to N\Delta(1232)$.  The same effect is seen  in
the pion-pion amplitude, where the inelasticity onset   in the
$s$-channel, $I=0$, starts at the energy much higher than the
four-pion threshold.  Presently there is quite a lot of evidences that
this is a general feature of the processes at low and intermediate
energies:  inelasticities are mainly related to the transition of
interacting hadron into an excited state, while the prompt production
of new quark-antiquark pairs is suppressed. This property being
observed several decades ago (see, for example, \cite{akk} and
references therein)
is discussed starting from the early 60's; it is a
foundation of duality ideas and models, such as Veneziano's  model
\cite{venez}.

 The direct way to perform a relativistic description
 of nucleon-nucleon interactions is to use the
invariant Feynman technique that leads to the Bethe -- Salpeter
equation.
However, the direct application of the Bethe -- Salpeter equation to
the nucleon--nucleon interaction faces problems related
to mass-off-shell amplitudes and
 meson production processes.  Mass-off-shell phenomena are usually
taken into account by the introduction of  form factors, that
undoubtedly affords an ambiguity in the quantitative representation of
the process.

Another problem arises in the description of meson
production within the Bethe-Salpeter equation.
Indeed, the solution of this
equation is represented as a sum of ladder diagrams
of the type shown  in Fig. 2.  The cutting of ladder diagrams defines
inelasticities of the amplitude; an example of such a cutting is shown
in Fig. 2 by dashed line, it corresponds to the transition $NN \to NN +
meson$.  The problem is that the forces in the deuteron channel related
to the pion exchange are rather significant.  Therefore, due to the
cuttings of Fig.2 -- type, the Bethe --- Salpeter equation leads to
significant inelasticities which are
switched on at invariant energies $\sqrt{s}=2m+m_{\pi},2m+2m_{\pi},$
etc.  To suppress the early inelasticities,
in line with the suppression of genuine inelastic processes,
 and at the same time to
hold a non-small interaction  related to pion-exchange forces, one
 should take into account  other more complicated
diagrams, in order 1) to cancel the  right-hand $s$-channel
singularities with the help of a
 destructive interference, and 2) to keep the
neighbouring $t$-channel singularities (or left-hand ones) that respond
to the non-small meson--exchange
forces.

Moreover, an exact summation of
ladder diagrams is hardly realizable in the realistic case, thus
 necessary simplifications being induced. One of them consists in the
consideration of  a potential-type interaction.  Another one is
an actual replacement of the constituent (nucleon) propagator
 by its residue $(k^2-m^2)^{-1} \to 2\pi
i\theta(k_0)\delta(k^2-m^2)$ \cite{gross}. However,
this latter procedure
leads to  anti-casual singularities of the amplitude; hence, one should
take care  these singularities be far from the region the approach
pretends to describe.

A correct handling of the $s$- and $t$-channel
singularities of the amplitude may be performed comparatively easily
 in the framework of
the dispersive N/D method \cite{chewmand}, where
contributions of the left-hand singularities  due to the
$t$- and $u$-channel exchanges are independent of the
right-hand singularities which are  related to the $s$-channel elastic
and inelastic processes.  An advantage of this technique is its
relativistic invariance, together with an absence of a problem of
  the   off-shell amplitude definition. For all
intermediate particles, the equality $k^2=m^2$ is fulfilled, thus
allowing us to describe interactions using mass-on-shell amplitudes.
 Moreover, in the dispersion relation technique there are no
principal problems with the description of inelasticities related both
to the excited baryons and prompt meson production, because all
$s$-channel intermediate states are
under control. This means that, when using dispersive
integrals, one can take into account certain set of the nearest
singularities in due course.

It is also important that the
dispersion relation
technique allows one to determine the relativistic wave function
(or vertex function) of the
composite (deuteron-like) system.
The description of two-particle composite systems within the
framework of the dispersive N/D method was developed in ref.
\cite{akms}}, where a Lorentz-covariant amplitude for
  the two-particle composite system interacting with an external
vector field has been written, and ambiguities related to the
dispersion relation subtractions have been eliminated basing on the
gauge invariance and analyticity requirements.  Therefore, while in the
Feynman diagrammatic technique the gauge invariance is automatically
fulfilled and the interaction amplitude obeys the Ward identity, in the
dispersion relation
technique these general properties of the theory should be
considered as supplementary constraints. This is a necessary payment for
treating independently left- and right-hand singularities.

In ref. \cite{akms} the  dispersion
relation analysis
 of the $pn$ scattering amplitude in the $^3S_1 -^3D_1$
channels has been performed. Basing on the phase shift  data,
the vertices of the $pn$ scattering in this channel have been
constructed together with the deuteron vertex functions which  are
relativistic analogues of the deuteron wave functions.  The use of
deuteron wave function obtained in such a way allows to describe well
the deuteron form factors up to $Q^2=2$ GeV$^2/$c$^2$ as well as
 deuteron binding energy and magnetic and quadrupole moments.  A
successful description of deuteron properties has been obtained taking
 properly into account the relativistic effects in the $pn$-component,
 without including non-nucleonic degrees of freedom into deuteron wave
 function.

The application of the
dispersion relation technique to the description of a deuteron
 has a long history (see refs. \cite{mandel}):
 at the early stage,
the deuteron form factor has been considered  using the
$t$-channel dispersion relation, thus bringing  the
 $t$-channel anomalous singularity  into consideration.
The recent deuteron description in terms of dispersive integrals
\cite{akms} is based on the double dispersion representation in the
deuteron--mass channels, and it
 does not require special treatment of anomalous
singularities. In this point the technique of ref.
 \cite{akms} is analogous  to the non-relativistic
wave function technique. The detailed consideration of the
composite-particle wave
functions in the dispersive technique and its relation to the technique
of light-cone variables may be found in refs. \cite{akms}, \cite{ammp}.

 Successful description of the elastic
deuteron form factors within the dispersion relation technique, gives
rise to the problem of the application of this technique to the other
two-nucleon processes, in particular to the deuteron
photodisintegration.

  The  reaction $\gamma d \to pn$ was  studied
during several decades
\cite{grosshol}--\cite{levchuk} in the framework of other techniques,
mainly in the quantum mechanical  and Feynman diagram approaches.
These investigations clarify many  features of the
reaction, among which --- an important role of the
meson exchange currents \cite{Laget} (typical diagrams are
shown in Fig. 3).

The analysis of the deuteron photodisintegration in the framework
of the dispersion relation technique
 have been carried out in \cite{alex1}-\cite{alex2}
starting from  small energies, where the contribution of inelastic
processes is negligible,  with  a subsequent increase of the
energy. In the present paper we expand the analysis onto
  the  region of the photon
energy $E_\gamma \simeq 250-400$ MeV, where the production of the
lightest baryons, $\Delta(1232)$ and $N^*(1400)$, is important.

The investigation of the reaction $\gamma d \to pn$ in the dispersive
integral technique complements the investigations carried out
within other approaches. This is related to the fact
that similar processes from the point of view of their diagrammatic
presentation may be treated in different approaches in their own ways.
Such an example is provided by meson exchange currents, for in the
dispersive technique they come from the contribution of
anomalous $t$- and $u$-singularities to the amplitude $\gamma d \to
pn$.

The paper is organized as follows.

In Chapter 2 the
nucleon-nucleon amplitude is analysed in the framework of the
dispersion  relation
technique in the energy range up to $T=1$ GeV. The aim of
this analysis is to restore nucleon-nucleon  amplitude
vertex functions, which
play the role of forces of the quantum mechanics approach. These
vertices are  used in the next Chapter for the calculation of the final
state interaction of nucleons in the deuteron photodisintegration. In
Section 2.1 the operator expansion of nucleon-nucleon amplitude in
partial waves is done, together with  the method of constructing
covariant partial wave operators.
In Sections 2.2 and 2.3, the dispersion relation N/D method is
described: it allows  to separate left- and right-hand
singularities or, in other words, to single out the
interaction forces from the
 production processes. Here the equation for the partial amplitude
is written for the one-channel case when inelastic processes are
small, and it is also shown how to present the amplitude using the
diagrammatic language.  In Section 2.4, basing on the phase shift
analysis, the  $N/D$ amplitude is restored in the channels
 $^1P_1$ and $^3D_2$, where inelasticity is small.
  In Section 2.5,  the diagrammatic technique
is generalized for the two-channel case. This allows us to include
inelasticity into the nucleon-nucleon amplitude and  expand our
analysis on other waves.  The channels $^1D_2$, $^3P_2$ and $^3F_3$
with the inelasticity  related to the production of $\Delta(1232)$
are considered here.  Basing on the phase shift data in the channels
$NN$ and $N\Delta$, the amplitudes are restored for the coupled
channels $^1D_2(NN)-^5S_2(N\Delta)$, $^3P_2(NN)-^5P_2(N\Delta)$ and
$^3F_3(NN)-^5P_3(N\Delta)$.  In Section 2.6 we briefly discuss the
waves $^3S_1$ and $^3D_1$: these waves were investigated in detail in
ref. \cite{akms}.
In Section 2.7  the structure of
inelasticities in the non-resonance channels $^1S_0$, $^3P_0$ and
$^3P_1$ is studied:  several mechanisms are considered, such as the
production of $\Delta(1232)$, $N^*(1440)$ or the $S$-wave $\pi N$ pair.
The technique for the calculation of three-particle loop diagrams which
are necessary for the description of inelastic processes is discussed
in Appendix A, there also the $N$-function parameters are given.

In Chapter 3, within the dispersion relation technique, the
amplitude $\gamma d \to pn$ is calculated up to the incident photon
energy  $E_{\gamma}=400$ MeV. The photodisintegration
amplitude is taken as a sum of diagrams shown in Fig. 1.
 Dashed blocks in Fig. 1b-f
correspond to the nucleon final state interaction, which includes
 elastic--scattering diagrams of Fig. 4a--type and
intermediate-state particle production  processes of Fig. 4b,c--type.
 In Section 3.1 the pole
diagram  of Fig. 1a is written in terms of dispersion relation
 expression in the deuteron channel. In Section 3.2 final
state interaction is calculated for the channels $^1S_0$,
$^3S_1-^3D_1$, $^1P_1$, $^3P_0$, $^3P_1$, $^3P_2$, $^3D_2$,
 $^1D_2$ and $^3F_3$ taking into account the diagrams of Fig. 4a-type
only (i.e.  neglecting inelastic rescatterings). Such an approximation
allows one to describe the
photodisintegration amplitude up to $E_{\gamma}=10$ MeV.  In
Section 3.3 inelasticities related to the production
of $\Delta(1232)$ are included into the final--state channels
  $^3P_2$, $^1D_2$ and $^3F_3$. Two mechanisms contribute
to the intermediate $\Delta$-production: the $\Delta$-production in the
final-state nucleon-nucleon interaction block (Fig. 4b) and  the prompt
$\Delta$-production  by photon (Fig. 1e).
 In Section 3.4 the contribution of inelasticities into $\gamma d \to
pn$ is calculated for nucleon--nucleon waves
$^1S_0$, $^3P_0$ and $^3P_1$.
It is shown that the process $\gamma d \to pn \to X \to pn$, where
$X$ is one of inelastic states $(N\pi)_S N$, $N^*(1400)N$ or $N\Delta$,
dominates in these waves. These transitions are important at $E_\gamma=
50-100$ MeV, while  the transition $\gamma d \to N\Delta \to pn$
in the waves $^1D_2$, $^3P_2$ and $^3F_3$, Fig. 1e, is important at
$E_\gamma \simeq 300-400$ MeV.

The detailed discussion of the
results is presented in Chapter 4.

\section{Dispersion Relation Representation of
Nucleon-Nucleon Scattering Amplitude }

\subsection{  Nucleon-
Nucleon Vertex Operators for Partial Amplitudes}

Consider the structure of multipole operators for the $NN$
partial scattering amplitudes. There are two ways for the presentation
of the amplitude expanded in multipoles. Standard
matrix element for the transition $NN\rightarrow NN$ (see Fig. 5a)
is as follows:
$$(\bar{\Psi}(p'_1)Q^b\Psi(p_1))\,
(\bar{\Psi}(p'_2)\tilde{Q}^b\Psi(p_2))\,,\eqno{(2.1)}$$
where $Q^b$ and $\tilde{Q}^b$  are multipole operators.
However, in the dispersive technique it is
more convenient to use charge-conjugated fields:
$$\Psi_c(-p'_2)=-\bar \Psi(p'_2)C\,,\qquad
\bar{\Psi}_c(-p_2)=C{\Psi}(p_2)\,,\eqno{(2.2)}$$
 where $C$ is  charge-conjugation operator. Using the Fierz
transformation \cite{fierz},
$\Psi(p_1) \leftrightarrow \Psi_c(p_2')$, one
gets
$$(\bar{\Psi}(p_1')\tilde Q^b\Psi_c(-p'_2))\,
(\bar{\Psi}_c(-p_2)Q^b\Psi(p_1))\,,\eqno{(2.3)}$$
where $Q^b$ and $\tilde{Q}^b$ is another set of operators
corresponding to the $^{2S+1}L_J$  $NN$ state.
The whole amplitude, $A(s,t)$,  depends on $s=(p_1+p_2)^2$  and
$t=(p_1-p'_1)^2$ and can be represented as a sum of partial
amplitudes, $A^b(s)$, as follows:
$$A(s,t)=\sum_{b}
(\bar{\Psi}(p'_1){I}_{ij}^b\tilde{Q}^b_{\mu\nu\cdots}
\Psi_c(-p'_2))
(\bar{\Psi}_c(-p_2){I}_{ij}^bQ^b_{\mu\nu\cdots}\Psi(p_1))
A^b(s)\,.\eqno{(2.4)}$$
Here $Q^b_{\mu\nu\cdots}$ and $\tilde{Q}^b_{\mu\nu\cdots}$ are
left and  right operators of the partial amplitude,
 $\mu,\nu...$ are indices  related to the
to the state $^{2S+1}L_J$, and
${I}_{ij}^b$ is  isotopic operator. Summation in eq. (2.4)
is carried out over the whole set of operators $Q^b_{\mu\nu\cdots}$.

Consider the structure of the $Q^b_{\mu\nu\cdots}$ operator.
For the   spin-$\frac 12$ particles,
these operators are constructed using metrical tensor
$g_{\mu\nu}$, antisymmetrical tensor
$\varepsilon_{\mu\nu\alpha\beta},$ $\gamma$-matrices
and momenta $k_\mu=\frac12 (p_1-p_2)_\mu$ and
$P_\mu=(p_1+p_2)_\mu$. Let us introduce the tensor
$T_{\mu'\nu'\cdots,\mu\nu\cdots}$ related to the two-nucleon loop
diagram:
$$\langle Sp[-(\hat{p}_1+m)\tilde{Q}^b_{\mu\nu\cdots}(-\hat{p}_2+m)
Q^a_{\mu'\nu'\cdots}]\rangle
=\delta^{ab}T_{\mu'\nu'\cdots,\mu\nu\cdots}
\rho(s).\eqno{(2.5)}$$
 Here $\delta ^{ab}$ is the Kronecker tensor, the
brackets $\langle\cdots\rangle$ denote the integraging over
two-particle phase space,
$d\Phi(p_1,p_2)$:  $$d\Phi(p_1,p_2)=\frac12
\, \frac{d^4p_1d^4p_2}{(2\pi)^6}(2\pi)^4\delta^4(P-p_1-p_2)
\delta(m^2-p_1^2)\delta(m^2-p_2^2),\eqno{(2.6)}$$
where
$$\int d\Phi(p_1,p_2)=\frac{1}{16\pi}\sqrt{\frac{s-4m^2}{s}}
\equiv\rho(s)\,.\eqno{(2.7)}$$
We impose the
     following normalization constraint:
$$T_{\alpha\beta\cdots,\mu\nu\cdots}T_{\mu'\nu'\cdots,
\alpha\beta\cdots}=T_{\mu'\nu'\cdots,\mu\nu\cdots}\,.
\eqno{(2.8)}$$
For the states with total angular momenta $J=0,1,2$,
the $T$-operators are equal to:
$$ \begin{array}{cc}
J=0:& T=1\,,\\
J=1:& T_{\mu',\mu}=g_{\mu'\mu}^\bot\,,\\
J=2:&
T_{\mu'\nu',\mu\nu}=\frac 12 \left(g_{\mu\mu'}^\bot
g_{\nu\nu'}^\bot+g_{\mu\nu'}^\bot g_{\nu\mu'}^\bot-\frac 23
g_{\mu\nu}^\bot g_{\mu'\nu'}^\bot\right)\,,
\end{array}
\eqno{(2.9)}$$
where
$g_{\mu\nu}^\bot=g_{\mu\nu}-P_\mu P_\nu /s$.

With these requirements, we get the following set
of the  $pn$ vertex operators  for the  states
$^{2S+1}L_J$ with isotopic spins $I=0,1$:

$$ \begin{array}{ccc}
^1S_0(I=1):&Q=\frac{1}{\sqrt{2s}}\gamma_5,& \tilde{Q}=-Q,\\

^3P_0(I=1):& Q=\frac{1}{\sqrt{2(s-4m^2)}},& \tilde{Q}=Q,\\

^3S_1(I=0):& Q_\mu=\frac{1}{\sqrt{2s}}\Gamma_\mu, &
\tilde{Q}_\mu=-Q_\mu,\\

^3D_1(I=0):& Q_\mu=\frac{1}{2\sqrt{s}}
\left[4k_\mu\frac{m+\sqrt{s}}{s-4m^2}+\gamma_\mu^\bot\right], &
  \tilde{Q}_\mu=-Q_\mu,\\

^1P_1(I=0):& Q_\mu=\sqrt{\frac{3}{2s(s-4m^2)}}\gamma_5k_\mu^\bot, &
\tilde{Q}_\mu=Q_\mu,\\

^3P_1(I=1):&
Q_\mu=\sqrt{\frac{3}{s^2(s-4m^2)}}\varepsilon_{\mu\alpha\beta\delta}
\gamma_\alpha P_\beta k_\delta, &
\tilde{Q}_\mu=-Q_\mu,\\

^3P_2(I=1):&
Q_{\mu\nu}=\sqrt{\frac{3}{2s(s-4m^2)}}\left[k_\mu^\bot\Gamma_\nu+
k_\nu^\bot\Gamma_\mu-\frac 23
(k^\bot\Gamma)g_{\mu\nu}^\bot\right], &
\tilde{Q}_{\mu\nu}=Q_{\mu\nu},\\

^1D_2(I=1):&
Q_{\mu\nu}=\frac{\sqrt{60}}{\sqrt{s}(s-4m^2)}\gamma_5
\left[k_\nu^\bot k_\mu^\bot-\frac{k^2}{3}g_{\mu\nu}^\bot\right],&
\tilde{Q}_{\mu\nu}=-Q_{\mu\nu},\\

^3D_2(I=0):&
Q_{\mu\nu}=\frac{\sqrt{10}}{s(s-4m^2)}\left[k_\nu^\bot
\varepsilon_{\mu\alpha\beta\delta}\gamma_\alpha P_\beta
k_\delta
+k_\mu^\bot\varepsilon_{\nu\alpha\beta\delta}\gamma_\alpha
P_\beta k_\delta\right],&
\tilde{Q}_{\mu\nu}=Q_{\mu\nu},\\

^3F_3(I=1):&
Q_{\mu\nu\chi}=\sqrt{\frac{3}
{70s^2(s-4m^2)^3}}
\end{array}$$
$$\times\left[K_{\mu\nu}^a\varepsilon_{abc\chi}\gamma_bP_c+
K_{\nu\chi}^a\varepsilon_{abc\mu}\gamma_bP_c+
K_{\chi\mu}^a\varepsilon_{abc\nu}\gamma_bP_c\right],\;
\tilde{Q}_{\mu\nu\chi}=Q_{\mu\nu\chi}\,.\eqno{(2.10)}$$
Here
$k_\mu^\bot=k_\mu-(Pk)P_\mu/s,\;\;\;$
$\gamma_\mu^\bot=\gamma_\mu-\hat{P}P_\mu/s,\;\;\;$
$\Gamma_\mu=\gamma_\mu^\bot-2k_\mu^\bot/(2m+\sqrt{s})\;\;\;$ and\\
$K_{\mu\nu\chi}=k_\mu k_\nu k_\chi-\frac{k^2}{5}
(k_\mu g_{\nu\chi}^\bot+k_\nu g_{\chi\mu}^\bot+k_\chi
g_{\mu\nu}^\bot).$ Let us stress that in (2.10) we have introduced
left and right operators because
 the trace (2.5) may be the negative in the technique used:
this leads to different  left and right operators.

It should be noted that $\Psi_c$  and $\Psi$ have  opposite
parities, so $\bar{\Psi}_c\gamma_5\Psi$ is a scalar and
$\bar{\Psi}_c\Psi$ is a pseudoscalar.

For the singlet and triplet isotopic states, the isotopic operators
 are:
$$I=0:\;\;\;I_{ij}=\frac{\delta_{ij}}{\sqrt{2}},
\qquad
I=1:\;\;\;\vec {I}_{ij}=\frac{\vec {\tau}_{ij}}{\sqrt{2}},\eqno{(2.11)}$$
where $\vec {\tau}_{ij}$ are  Pauli
matrices.

Consider as an example the construction of
the operator $Q_{\mu\nu}$ for the $^3P_2$ state:
the total momentum $J=2$, so this operator is  traceless symmetrical
tensor of the second rank
constructed of  $k^{\perp}_{\mu}$ (the $P$ wave state) and
$\Gamma_{\nu}$ (the spin-triplet state). Therefore,
the operator for the $^3P_2$ state is proportional to
$O_{\mu\nu}=
k^{\perp}_{\mu}\Gamma_{\nu}+k^{\perp}_{\nu}\Gamma_{\mu}-\frac 23
(k^{\perp}\Gamma) g^{\perp}_{\mu\nu}$, where
$Q_{\mu\nu}=N(s)O_{\mu\nu}$ and $N(s)$ is normalization constant.
We have:
$$\langle Sp[-(\hat{p}_1+m)\tilde{O}_{\mu\nu}
(-\hat{p}_2+m)O_{\mu'\nu'}]\rangle
=\frac 23 s(s-4m^2)T_{\mu \nu,\mu' \nu'}\rho(s). \eqno{(2.12)}$$
Therefore, the operator for the $^3P_2$ state is:
$$Q_{\mu\nu} =\sqrt{\frac{3}{2s(s-4m^2)}}O_{\mu\nu} .\eqno{(2.13)}$$
Likewise, one can construct vertex operators for all partial
amplitudes.

The partial amplitude $A^b(s)$ for a fixed state is obtained by
projecting the whole amplitude  $A(s,t)$ given by (2.4)
on this state. Corresponding conjugate
operator is as follows:
$$(\bar{\Psi}(p_1)Q^b_{\mu\nu...}\Psi_c(-p_2))\,
(\bar{\Psi}_c(-p'_2)\tilde{Q}^b_{\mu'\nu'...}\Psi(p'_1)).\eqno{(2.14)}$$
Thus, we have:
$$
4T_{\mu\nu...\mu'\nu'...}\rho(s)A^b(s)\rho(s)=
\int d\Phi(p_1,p_2)d\Phi(p'_1,p'_2)$$
$$\times
(\bar{\Psi}(p_1)Q^b_{\mu\nu...}\Psi_c(-p_2))\,A(s,t)\,
(\bar{\Psi}_c(-p'_2)\tilde{Q}^b_{\mu'\nu'...}\Psi(p'_1)).
\eqno{(2.15)}$$
After performing azimuthal-angle integration,
 the following expression for $A^b(s)$
is obtained:
$$
4T_{\mu\nu...\mu'\nu'...}A^b(s)=
\int _{-1}^1 \frac {dz}{2}
(\bar{\Psi}(p_1){Q}^b_{\mu\nu...}\Psi_c(-p_2))A(s,t)
(\bar{\Psi}_c(-p'_2)\tilde {Q}^b_{\mu'\nu'...}\Psi(p'_1)),
\eqno{(2.16)}$$
where $z$ is the cosine of the angle between $\vec p_1$ and $\vec p'_1$
in the center-of-mass system of
particles 1 and 2: $z=1+2t/(s-4m^2)$.

\subsection{Left-Hand and Right-Hand Singularities
and Unitarity of Partial Amplitudes}

In this Section we remind the analytic structure  of the
 amplitude when nucleons interact by pion exchanges
($m_\pi$ is pion mass). In this case the amplitude has
a pole in the $t$-channel at $t=m_\pi^2$,  as well as
threshold singularities at
  $t=(nm_\pi)^2$ $(n=2,3,... ) $. In the $s$-channel the
amplitude has a singularity at  $s=4m^2$ related to nucleon
rescattering and branching points  $s=(2m+nm_\pi)^2$ related  to
inelastic channels.  For the deuteron channel a bound state with the
mass $M$ exists: the amplitude has a pole singuliarity at $s=M^2$.  The
$s$-channel partial amplitudes have  the same  $s$-channel
singuliarities (right-hand ones) that the whole  amplitude, $A(s,t)$,
has. Besides, partial amplitudes have left-hand singularities which
correspond to the $t$- and $u$-channel singuliarities of $A(s,t)$ with
branching points $s=4m^2-m_\pi^2n^2$  --- see
Fig. 6.  Neglecting inelasticities means that we neglect all  right-hand
cuts, except for the first one with the branching point at $s=4m^2$.

Consider the unitarity of the partial amplitude in the
physical region (the upper edge of
 the elastic cut). One has $SS^+=1$ with
$$
S={\bf 1}+i(2\pi)^4\delta^4(p_1+p_2-p'_1-p'_2) A(s,t).
\eqno{(2.17)}$$
 Using the decomposition of the amplitude (2.4) and equation
(2.5), we get the unitarity relation for the partial
amplitude:
$$
Im\, A^b(s)=\rho(s)|A^b(s)|^2.\eqno{(2.18)}$$

\subsection{ One-Channel Scattering in the Diagrammatic Dispersion
Relation Technique}

 According to the $N/D$ method \cite{chewmand}, the partial
amplitude $A(s)$ can be represented as a ratio of  two functions
(here and below we omit the index $b$ in the partial $NN$ state):
$$
A(s)=\frac{N(s)}{D(s)},
\eqno{(2.19)}$$
with $N(s)$ containing  left-hand sinquliarities and $D(s)$ containing
 right-hand ones only.  It
follows from the unitarity condition (2.18) that
$$Im\, D(s)=-\rho(s)N(s).
\eqno{(2.20)}$$
The used normalization  for $D(s)$  reads: $D(s)\rightarrow 1$ at
$s\rightarrow\infty$. It is also assumed that there are no  CDD-poles
\cite{CDD}
in the amplitude. In this way one gets:
$$
D(s)=1-\int_{4m^2}^{\infty}\frac{ds'}{\pi}\frac{\rho(s')N(s')}{s'-s}
=1-B(s).
\eqno{(2.21)}$$
One can restore the $N$-function
using experimental data, thus solving the problem of finding
an amplitude with correct analytic properties in the vicinity of the
point $s=4m^2$. However, to describe the
nucleon-nucleon system (say, deuteron)  which
interacts with an external field, it is convenient to introduce
 vertices, which, in the case of a deuteron, play the role of  wave
functions  \cite{anisbugg}. In the
simplest case, the vertex function may be defined as follows:
$$
G(s)=\sqrt{N(s)}\,.
\eqno{(2.22)}$$
Then $A(s)$  can be
written as a sum of  dispersive diagrams shown in Fig. 7a.
  $$A(s)=G(s)
(1+B(s)+B(s)^2+\cdots) G(s),\eqno{(2.23)}$$
where $B(s)$ corresponds to the one-loop diagram with
the vertex $G(s)$.

Consider the one-loop dispersive diagram and its connection with
the Feynman one-loop diagram. The Feynman integral reads:
$$
A_F=\int \frac{d^4k}{(2\pi)^4i}\frac
{a(k^2_1, (P-k_1)^2;k^2,(P-k)^2)a(k^2, (P-k)^2;k^2_2,(P-k_2)^2)}
{(m^2-k^2)(m^2-(P-k)^2)}.
\eqno{(2.24)}$$
Here $a(k^2_1, (P-k_1)^2;k^2,(P-k)^2)$ is an irreducible block, without
two-particle intermediate states. The receipt for the transition from
Feynman integral to the dispersive one is the following. We need:
\\1) To neglect right-hand multiparticle singularities in the block
$a$;
\\2) To factorize the block $a$:
$$
a(k^2_1, (P-k_1)^2;k^2,(P-k)^2) \to G(k^2_1,
(P-k_1)^2)G(k^2,(P-k)^2);
\eqno{(2.25)}$$
\\3) To calculate  imaginary part of the one-loop Feynman diagram,
considering  the intermediate state with the energy $\sqrt{s}$
 as a real one $(s>4m^2)$:
$$
 \frac{d^4k}{(2\pi)^4i} \frac1{(m^2-k^2)(m^2-(P-k)^2)}$$
$$ \to
\frac{d^4k}{(2\pi)^4i}\frac1{2i} (2\pi i)^2 \theta(k_0)\delta(m^2-k^2)
\theta(P_0-k_0)\delta(m^2-(P-k)^2)=\rho(s);
\eqno{(2.26)}$$
\\4) To restore the $B$-function using dispersion relation integral
over the imaginary part related to the elastic cut, $disc\,A_F$:
$$\int ^\infty _{4m^2} \frac{ds'}{\pi} \frac{\rho(s')}{s'-s}disc\,
A_F(s')=\int ^\infty _{4m^2} \frac{ds'}{\pi} \frac{G^2(s')\rho(s')}
{s'-s}.\eqno{(2.27)}$$

Finally, we get $G(s)B(s)G(s)$.

The above consideration demonstrates the applicability limit for
eq. (2.24): the basic assumption is a factorization of the block
$a$.

A general method to perform a factorization of the $t$- and
$u$-channel meson exchange interactions has been developed in
\cite{anisbugg}.
For the description of a composite system in  general case, one
should introduce several vertex functions $G_i$ related to a separable
interaction $V_i$: $N_i\rightarrow V_i=G_iG^i$, where the
left vertex function,  $G_i$, and the right one, $G^i$, may be
different. For further consideration, it
 is convenient to introduce the amplitudes $a_i$
which do not contain the right vertices $G^i$. The whole amplitude and
  auxiliary amplitudes $a_i$ are related as:
  $$
A=\sum_{i}a_iG^i.
\eqno{(2.28)}$$
The amplitudes $a_i$ obey  a set of linear equations:
  $$
a_i=\sum_{j}a_jB_i^j+G_j\delta_{ij}\,,
\eqno{(2.29)}$$
where (see Fig. 7b)
$$B_i^j(s)=\int_{4m^2}^{\infty}\frac{ds'}{\pi}
\frac{G^j(s')\rho(s')G_i(s')}{s'-s}\,.
\eqno{(2.30)}$$
Rewriting (2.29) in a matrix form, one has
$$\hat{a}=\hat{B}\hat{a}+\hat{g}\,,
\eqno{(2.31)}$$
where
$$\hat{a}=\left[\begin{array}{c}
a_1\\a_2\\.\end{array}\right]\,,\quad
\hat{g}=\left[\begin{array}{c}
G^1\\G^2\\.\end{array}\right]\,,\quad
\hat{B}=\left[\begin{array}{ccc}
B_1^1&B_1^2&.\\
B_2^1&B_2^2&.\\
.&.&.\end{array}\right]\,.
\eqno{(2.32)}$$
The final expression for the partial amplitude reads:
$$
A=\hat{g}^T(I-B)^{-1}\hat{g}\,,\eqno{(2.33)}$$
where
$$\hat{g}^T=[G_1,G_2,\cdots]\,.\eqno{(2.34)}$$

The amplitude obtained in such a way is unitary; it may be derived
from the unitarity condition directly.

Now let us explain how the scattering amplitude discussed above
may be written in terms of the energy-off-shell Bethe-Salpeter equation.
We need to introduce the energy-off-shell scattering amplitude:
$$A(s,s_1)=\sum _i a_i(s)G^i(s_1).
\eqno{(2.35)}$$
Using equation (2.31), one can write this amplitude as
$$
A(s,s_1)=
\int ^\infty _{4m^2} \frac{ds'}{\pi} \frac{\rho(s')}{s'-s}
 A(s,s')V(s',s_1)+V(s,s_1).
\eqno{(2.36)}$$
This is a dispersive Bethe-Salpeter equation for $A(s,s_1)$, where the
 effective interaction is
$$V(s,s_1)= \sum _i G_i(s)G^i(s_1).
\eqno{(2.37)}$$

The suggested diagrammatic technique is a modification of
the standard $N/D$ method: this technique is suitable for the
description of an interaction with an external field.

\subsection{One-Channel Approach for   the Waves $^1P_1$ and  $^3D_2$}

For   the waves $^1P_1$ and  $^3D_2$ the inelasticity is small, and
corresponding amplitudes can be described in the one-channel
approximation.
We  restore vertex functions of the
$pn$ scattering, relying upon  the phase shift analysis data
\cite{arndt}.
The $G$-function is determined by the left-hand
 singularities, therefore it is represented as a dispersive
integral along the left-hand cuts:
$$
G(s)=\int_{-\infty}^{s_L}\frac{ds'}{\pi}\frac{disc\,G(s')}
{s-s'},
\eqno{(2.38)}$$
where $s_L=4m^2-m_\pi^2$ is the location of the nearest pion
branching point. Below, to simplify a cumbersome calculation, we
replace the integartion in (2.38) by the summation:
  $$
G_i(s)=\sum_{j=1}^{6}\frac{\gamma_j^i}{s-s_j^i}\,,\qquad
s_j^i=s_0^i-h^i(j-1), \eqno{(2.39)}$$
with the poles $s_j^i$   placed
 on the left from $s_L$.

  The values of parameters  $\gamma_j^i$,
 $h^i$ and $s_0^i$ obtained by  fitting   the phase
 shift data are discoursed in Appendix A.
The results of the phase shift fit together
  with the experimental data are demonstrated in Fig. 8.

\subsection {Two--Channel Scattering, $NN$ and $N\Delta$:
\newline the Waves
 $^1D_2$, $^3P_2$ and $^3F_3$}

In this Section we consider the inelasticity in the nucleon-nucleon
amplitude due to the production of the isobar $\Delta(1232)$.
The nucleon-nucleon amplitude is treated as a two-channel one (the
channels $NN$ and $N\Delta$). We analyse here the coupled channels:
$^1D_2(NN)-^5S_2(N\Delta)$, $^3P_2(NN)-^5P_2(N\Delta)$ and
$^3F_3(NN)-^5P_3(N\Delta)$.

Matrix element for the $NN\rightarrow
N\Delta$ transition (see Fig. 5b for the momentum notation) is written
in the form
$$ (\bar{\Psi}^\mu(p'_1) \tilde
{Q}^a_{\alpha\beta,\mu}T_0\Psi_c(-p'_2))
(\bar{\Psi}_c(-p_2)Q^b_{\alpha\beta}{I}\Psi(p_1))A^{ab}(s).
\eqno{(2.40)}$$
Here $\tilde Q^a$ and $Q^b$  are  vertex operators for the states
$a$ and $b$. $T_\lambda$ and $I$
are isotopic operators in the  $N\Delta$ and $NN$ channels,
correspondingly.  $A_{ab}(s)$ is the partial amplitude for the
transition $ state\,b \to state\, a$.  $\Psi_\mu$ is the wave function
for the spin-$\frac32$ particle;  $\Psi_\mu$ obeys the  equations:
$$p_\mu\Psi_\mu(p)=0, \qquad \gamma_\mu\Psi_\mu(p)=0.\eqno{(2.41)}$$
  For the $N\Delta$ states, vertex operators are as follows:
$$
 ^5S_2(N\Delta):\;\;\;
\tilde Q_{\alpha\beta,\mu}=
\frac{1}{\sqrt{\Omega_1}}\left[\tilde{\Gamma}_\alpha
g_{\beta\mu}^\bot+\tilde{\Gamma}_\beta g_{\alpha\mu}^\bot-\frac 23
g_{\alpha\beta}^\bot\tilde{\Gamma}_\mu\right]\equiv
S_{\alpha\beta,\mu},$$

$$^5P_2(N\Delta):\;\;\;
\tilde Q_{\alpha\beta,\mu}=\frac{1}{\sqrt{\Omega_2}}\left[\varepsilon_{abc\alpha}
k_a^tP_bS_{c\beta,\mu}+\varepsilon_{abc\beta}k_a^tP_bS_{c\alpha,\mu}
\right],$$

$$^5P_3(N\Delta):\;\;\;
\tilde Q_{\alpha\beta\gamma,\mu}=\frac{1}{\sqrt{\Omega_3}}\left[S_{\alpha\beta,\mu}
k_\gamma^t+S_{\beta\gamma,\mu}k_\alpha^t+S_{\gamma\alpha,\mu}k_\beta^t
\right.$$
$$\left.-\frac 25 k_\nu^t(S_{\alpha\nu,\mu}g_{\beta\gamma}^\bot+
S_{\beta\nu,\mu}g_{\gamma\alpha}^\bot+S_{\gamma\nu,\mu}g_{\alpha\beta}^\bot)
\right].
 \eqno{(2.42)}$$
Here $\Omega_i$ is a normalizing factor; the momenta are defined as
$P=p_1+p_2$, $k=\frac12 (p_1-p_2)$, $s=P^2$,
$k^t=\frac12 (p_1-p_2)-\frac{1}{2s}(m_\Delta^2- m^2)P$, where
 $m_\Delta$ is the $\Delta$-isobar mass; $k_\mu^\bot$,
 $\gamma_\mu^\bot$ and $g_{\alpha \beta}^\bot$ are introduced in
Section 2.1, while $\tilde \Gamma_\mu$ is defined as
$$\tilde{\Gamma}_\mu=\gamma_\mu^\bot-\frac{2k_\mu^\bot}
{m+m_\Delta+\sqrt{s}}\,.
 \eqno{(2.43)}$$
The $NN$
 vertex operators, $Q^b$, for $^1D_2$, $^3P_2$ and $^3F_3$
 states are given by  equation (2.10).

The isotopic operator $T_0$ for the states $\Delta ^+ n$ and
$\Delta ^0 p$ is defined as
$$T_0^+=\frac{1}{\sqrt{2}}\left[\begin{array}{cc}
0&0\\1&0\\0&1\\0&0\end{array}\right]\,,\quad
T_0=\frac{1}{\sqrt{2}}\left[\begin{array}{cccc}
0&1&0&0\\0&0&1&0\end{array}\right]\,.
 \eqno{(2.44)}$$

The partial  amplitude $A$ is $2\times 2$ matrix. It is related to
the $2\times 2$ $S$-matrix as follows:
 $$
 S=I+2i\sqrt{\rho} A
\sqrt{\rho}\,,
 \eqno{(2.45)}$$
where $\sqrt{\rho}$ is the matrix
of square roots of phase space factors:
  $$
 \sqrt{\rho}=\left[\begin{array}{cc}
\sqrt{\rho_{NN}} &0\\ 0&\sqrt{\rho_{N\Delta}}\end{array}\right]\,.
\eqno{(2.46)}$$

As before (see Section 2.3), the partial amplitude is represented in
the form:
$$A=\hat{g}^T(I-\hat {B})^{-1}\hat{g}\,,
 \eqno{(2.47)}$$
but with another structure of vertex operators:
$$
 \hat{g}=\left[\begin{array}{cc}
G^N&0\\G^t&0\\0&G^r\\0&G^\Delta\end{array}\right]\,,\qquad
\hat{g}^T=\left[\begin{array}{cccc}
G_N&0&G_t&0\\0&G_r&0&G_\Delta\end{array}\right],\eqno{(2.48)}$$
 and $B$-matrix
$$\hat{B}=\left[\begin{array}{cccc}
B_{NNN}&0&B_{NNt}&0\\B_{tNN}&0&B_{tNt}&0\\
0&B_{r\Delta r}&0&B_{r\Delta\Delta }\\
0&B_{\Delta\Delta r}&0&B_{\Delta\Delta\Delta}\end{array}\right],\;\;\;
B_{ijk}(s)=\int_{T_j}^{\infty}\frac{ds'}{\pi}\frac{G^i(s')\rho_j
(s')G_k(s')}{s-s'}\,.
\eqno{(2.49)}$$
Here $G_N G^N$, $G_\Delta G^\Delta$, $G_r G^t$ and $G_t G^r$
are  $N$-functions for
the transitions $NN\rightarrow NN$, $N\Delta\rightarrow N\Delta$,
$NN\rightarrow N\Delta$ and $N\Delta\rightarrow NN$, correspondingly
(see Fig. 5e), $T_1=4m^2$, $T_2=(2m+m_\pi)^2$. In our fit we put
$G^r=G_r$ and $G^t=G_t$; as it was in previous Section, the two-vertex
form is used for the transition $NN \to NN$: $G_N G^N=G_1 G^1 -G_2
G^2$ --- Fig.  5f.

For  vertex functions  the parametrization (2.39) is used.
Parameters for  $G$  are found  using  nucleon-nucleon and
nucleon-isobar phase shifts,
$\delta_{NN}$ and  $\delta_{N\Delta}$, and the inelasticity parameter
$\eta$ which are defined as
 $$
 S_{11}=\eta
e^{2i\delta_{NN}}\;\;, |S_{12}|=|S_{21}|=\sqrt{1-\eta^2}\;\;,
S_{22}=\eta e^{2i\delta_{N\Delta}}\,.
 $$
Here indices 1 and 2 refer to the channels $NN$ and $N\Delta$.
The description of the phase shift data
in the coupled channels $^1D_2-^5S_2$, $^3P_2-^5P_2$ and $^3F_3-^5P_3$
is shown in Fig. 9 together with the experimental data
\cite{phases}.

\subsection{Coupled channels $^3S_1$ and $^3D_1$}

In Fig. 8 the coupled channels $^3S_1-^3D_1$ are
 drawn, for which the inelasticity is small, though the connection of
channels is important and the two-channel approach must be applied
 here \cite{akms}.

\subsection{  Models for the Inelasticity in $^1S_0$ wave}

According to \cite{arndt},  the mechanism of
inelasticity  in the waves
  $^1S_0,$ $^3P_0$ and $^3P_1$ differs from the inelasticity
in the channels $^1D_2$, $^3P_2$ and $^3F_3$, and the amplitude has a
non-resonance behaviour.

Here we consider several models for inelasticity in the  $^1S_0$
wave. First, we suggest that the inelasticity comes due to the production
of $\Delta(1232)$. Second, we consider as a source of inelasticity
the production of $N^*_{11}(1440)$. This resonance is heavier
than  $\Delta(1232)$, but it has larger width, and the
transition amplitude $^1S_0(NN)-^1S_0(NN^*)$ is not suppressed
 near the threshold. Third,  the prompt pion production in the  $\pi
N$ $S$-wave is assumed. It should be noted that the  1st and 2nd
variants correspond to the production of the $\pi N$-state in
the $P$-wave .  \\
1) {\bf $\Delta(1232)$-isobar production}

Matrix element for the transition $^1S_0(NN)-^5D_0(N\Delta)$
is determined by  (2.40). The vertex operator for the $^5D_0$ wave
 is
$$^5D_0(N\Delta):\;\;\; Q_\mu=\frac{1}{\sqrt{\Omega}}
\left[\tilde{\Gamma}_\alpha g_{\beta\mu}^\bot+ \tilde{\Gamma}_\beta
g_{\alpha\mu}^\bot-\frac 23
g_{\alpha\beta}^\bot \tilde{\Gamma}_\mu\right] K_{\alpha\beta}\;\;,
\tilde Q_\mu=-Q_\mu, \eqno{(2.50)}$$
where  $\Omega$  is the
normalization factor and
$$K_{\alpha\beta}=k_\alpha^t k_\beta^t-\frac 13
k^{t2}g_{\alpha\beta}^\bot\,.\eqno{(2.51)}$$
The other quantities refer to equations (2.42) and
(2.43).\\
2) {\bf Production of $N^*(1440)$}

Matrix element for the transition $NN\rightarrow NN^*$ is  written
   in the form similar to  (2.40):
$$(\bar{\Psi}_{N^*}(p'_1)\tilde Q^aI^i\Psi_c(-p'_2))
(\bar{\Psi}_c(-p_2)Q^bI^i\Psi(p_1))A^{ab}(s)\,.\eqno{(2.52)}$$
The resonance $N^*(1440)$ has the same quantum numbers as a
nucleon, so  the vertex operator for the $NN^*$ state is:
$$^1S_0(NN^*):\;\;\;
Q=\frac{1}{\sqrt{\Omega}}\gamma_5\;\;\;,
 \tilde Q=-Q.\eqno{(2.53)}$$
3) {\bf Production of the $(\pi N)_S$-pair in the
$J^P=\frac12^-$ state}

  Since the lightest resonances in the $\pi N$ system,
with quantum numbers $J^P=\frac12^-$, are located far from
 the pion-nucleon
 threshold, we deal here with the case of the prompt non-resonance
pion production (Fig. 5c). Nevertheless, we approximate the $\pi N$
pair by a remote quasi-resonance $R$, with an
energy--dependent width. This enables us to operate with two-particle
intermediate states using the same technique as was used for
$\Delta(1232)$ and $N^*(1440)$.

Consider the vertex of the quasi-resonance decay $R \to \pi N$:
$\Gamma_\pi=c_\pi\vec {\tau}$.
The width of the quasi-resonance $R$, $\Gamma_R^w$, is defined by
its decay into pion and nucleon.  This allows us to relate the
quantities $c_\pi$ and $\Gamma_R^w$.  The resonance propagator  is
written as:  $$ \frac{(\hat p+M_R)}{M_R^2-p^2-B(p^2)}\,,
\eqno{(2.54)}$$
where $B(p^2)$ corresponds to the pion--nucleon loop diagram (Fig. 5d).
 Real part of the loop diagram re-determines
the mass of the quasi--resonance  $M_R$,
and  imaginary part of $B(p^2)$  provides  its width.
Let us rewrite the propagator (2.54) as a series:
$$
\frac{(\hat{p}+M_R)}{M_R^2-p^2
-iM_R\Gamma_R}=\frac{(\hat{p}+M_R)}{(M_R^2-p^2)} +
\frac{(\hat{p}+M_R)}{(M_R^2-p^2)}i\Gamma_R^wM_R\frac{1}{(M_R^2-p^2)}
+...\eqno{(2.55)}$$
where the first term corresponds to the propagator of a stable
particle and the second one to the propagator with the
inclusion of the  pion-nucleon loop diagram.
  The Feynman integral for the second term of the right-hand side
of eq. (2.55)  is equal to:
$$
\frac{(\hat{p}+M_R)}{M_R^2-p^2}
\int\frac{d^4k}{(2\pi)^4i}\frac{c_{\pi}^2 (\hat{k}+m)}
{(m^2-k^2)(m_\pi^2-(p-k)^2)}
\frac{(\hat{p}+M_R)}{M_R^2-p^2}\,.
\eqno{(2.56)}$$
The comparison of eqs. (2.55) and (2.56) provides the expression for
$\Gamma_R^w$ in terms of $c^2_\pi$. To this end,
let us decompose   $k$  in
external vectors $p$ and $k^\bot$, the latter being orthogonal to $p$:
$$k=\alpha p+k^\bot,\eqno{(2.57)}$$ where
$\alpha=(p^2+m^2-m_\pi^2)/2p^2$.
The term which is
proportinal to $k^\bot$ vanishes in eq. (2.56).
When comparing eqs. (2.55) and (2.56), it is necessary to calculate
imaginary part of the loop diagram and to neglect its real part; for
this purpose the following replacement should be done:
$$
\frac1{(m^2-k^2)(m_\pi^2-(p-k)^2}\rightarrow\frac1{2i}(2\pi i)^2
\delta(m^2-k^2)\theta(k_0)\delta(m_\pi^2-(p-k)^2)
\theta(p_0-k_0)\,.
\eqno{(2.58)}$$
Final expression for the width $\Gamma^w_R$ takes a form:
 $$
\Gamma^w_R=\frac{c_\pi^2}{8\pi
M_R\sqrt{s}}[(M_R+m)^2-m_\pi^2]
\left[\frac{(s+m^2-m_\pi^2)^2}{4s}-m^2\right]^{1/2}\,.
\eqno{(2.59)}$$
Vertex operator for the $NR$ system in the $^1S_0$ state is:
$$
^1S_0(NR):\;\;\;Q =\frac1{\sqrt{\Omega}}\,{\bf 1},\;\;\;\tilde Q =Q .
\eqno{(2.60)}$$

The description of the $NN$ amplitude
is performed within the framework of the two-channel amplitude,
where the first channel is the $NN$-state and  the second is
$N\Delta$, or $NN^*$, or $NN\pi$ states, depending
on the model we use. The amplitude is
restored  basing
on the phase shift analysis data of ref.  \cite{arndt}.
The results of fitting to the phase shift analysis data
for the $^1S_0$ wave     are shown in Fig. 10a,b.
All the three models work equally well in the description of the
experimental data on the elastic amplitude $NN \to NN$. However, they
provide different results for the amplitude  $NN \to NN\pi$.

\subsection{$NN^*(1400)$ Channel as a Source of Inelasticity
in $^3P_0$ and $^3P_1$ Waves}

The description of the $^3P_0$  and $^3P_1$ $NN$ waves   is performed
 assuming  the production of $N^*(1400)$ as a source of
inelasticity in the intermediate state.  The results of fitting
procedure for the $^3P_0$ and $^3P_1$ waves are shown in Fig. 10c-f.
Fitting parameters are given in    Appendix A.

\section{Deuteron Photodisintegration within  the Dispersion
Relation Technique  }

In this Section
 the deuteron photodisintegration amplitude is calculated basing on the
analysis performed in the previous Section.
This amplitude is represented as a sum of the pole (impulse
approximation) diagram (Fig. 1a), diagram with nucleon-nucleon
final state interaction  (Fig. 1b)
and diagrams with the photoproduction of
 $\Delta(1232)$, $N^*(1400)$ and pions in the intermediate state (Fig.
 1c-f). It should be stressed that all parameters which are
necessary for the calculation of these diagrams have been found in the
 Section 2, where nucleon -- nucleon interaction have been
studied; therefore the calculation of the $\gamma d \to pn$
amplitude is parameter-free.

\subsection{Pole Diagram}

Consider the amplitude $A^{pol}$, which
corresponds to the pole diagram of Fig. 11a, written within the
dispersion relation technique.  We start with the presentation of the
Feynman pole diagram:
$$ A^{pol}=\xi_\tau\varepsilon_\mu
M_{\tau\mu}^{pol}\,, \eqno{(3.1)}$$ where
$$
M_{\tau\mu}^{pol}=\bar{\Psi}(p_1)\Gamma_\mu\frac{\hat{p}+m}
{m^2-p^2}I^d\Gamma_\tau^d\Psi_c(-p_2)\,.
\eqno{(3.2)}$$
Here $\xi_\tau$ and $\varepsilon_\mu$ are  polarization vectors of
the deuteron and  photon which are orthogonal to the deuteron and photon,
$P\;\;(P^2=M^2)$ and $q\;\;(q^2=0)$, momenta:
 $$(\xi P)=0\,,\qquad
(\varepsilon q)=0\,.
\eqno{(3.3)}$$
The deuteron isotopic operator, $I^d(I=0)$, is
$$
I_{ij}^d=\frac{\delta_{ij}}{\sqrt{2}}\,.
\eqno{(3.4)}$$
Vertex functions for the transition  $deuteron \to
nucleons$, $\Gamma_\tau^d$, and   for the
photon-nucleon interaction, $\Gamma_\mu$, are of the form:
$$
\Gamma_\tau^d=\gamma_\tau\phi+(p-p_2)_\tau\varphi,\;\;\;
\Gamma_\mu=e(\gamma_\mu f_1+(p+p_1)_\mu f_2)\,,
\eqno{(3.5)}$$
where  the vertex
 functions  $f_1$ and $f_2$  are matrices in the isotopic space:
$$
f_i=(f_i^p+f_i^n){\bf 1}+(f_i^p-f_i^n){\bf \tau }_3\,.
\eqno{(3.6)}$$
Here ${\bf 1}$ and ${\bf \tau }_3$ are isotopic matrices  ${\bf
1}=diag(1,1)$ and ${\bf \tau }_3=diag(1,-1)$.  Indices $p$  and $n$
    stand for the photon-proton and photon-neutron form factors;
 $\phi$, $\varphi$  and $f_i^N(N=p,n)$ depend on invariant variables.

Consider the  transition from the Feynman pole amplitude (3.2) to
 the dispersion relation one. For this diagram the
 dispersion relation is written in the deuteron channel,  where
the pole diagram  contains  the two-particle threshold
 only, which  corresponds to the production of two nucleons in
 the intermediate state (in Fig. 11a the relevant
two-particle cut is
shown by  dashed line).

When going from the Feynman amplitude to the dispersive one, the
intermediate state vector
$\tilde {P}$ should be introduced,
see Fig. 11a: $\tilde {P}=p+p_2$, $\tilde {P}^2=\tilde s$. The
transition to the dispersive integral in the deuteron channel is
carried out at fixed total energy $s=(p_1+p_2)^2$.  Let us introduce
$\tilde q=-p+p_1$ which satisfies the equations:
$$ \tilde P+\tilde
q=p_1+p_2=P+q.\eqno{(3.7)} $$
For the calculation of kinematic
relations, it is suitable to use the deuteron rest frame, where
$$ q=(q_0, \vec q),\;\;\;
P=(M,  0),\;\;\; \tilde P=(\sqrt{\tilde s}, 0),\;\;\;
\tilde q=(M+q_0-\sqrt{\tilde s}, \vec q).
\eqno{(3.8)}
$$
One can see that $\tilde {q}^2=(p-p_1)^2 \ne q^2$. This inequality
is related to the energy non-conservation in the
dispersive integrals.
 The photon energy in the deuteron rest frame  $q_0$
is determined by $s$ as:
$$
M^2+2Mq_0=s.\eqno{(3.9)}
$$
We get from  (3.9) and (3.10):
$$
\tilde q^2=s+\tilde s-\frac{\sqrt{\tilde s}}{M}(s+M^2). \eqno{(3.10)}
$$

Calculation of the discontinuity  in the deuteron channel
is equivalent to the replacement of the propagator by the
$\delta$-function, considering  the  two-nucleon
intermediate state as a real one:
 $$
 \frac{1}{m^2-p^2}\rightarrow\frac{1}{2}(2\pi
 i)\delta(m^2-p^2)\theta(p_0)\,.
\eqno{(3.11)}$$
 Using the discontinuity across the two-nucleon
cut, we reconstruct the amplitude
$$
M_{\tau\mu}^{pol}=\int _{4m^2}^{\infty}
\frac{d\tilde s}{\tilde s-M^2}\delta(m^2-p^2)\theta(p_0)
\bar{\Psi}(p_1)\Gamma_\mu(\hat{p}+m)I^d\Gamma_\tau^d(\tilde s)
\Psi_c(-p_2)
 \,.
\eqno{(3.12)}$$
The substitution (3.11)
corresponds to  physical decay of the virtual deuteron with
 momentum $\tilde P$ into two nucleons with the
on-shell momenta $p$ and $p_2$.   Rewriting the $\delta$-function
argument in the form
$$
p^2-m^2=\tilde s-2\tilde P p_2, \eqno{(3.13)}$$
we elimitate the
 integration in (3.12):
$$
M_{\tau\mu}^{pol}=\frac{2}{\tilde{s}-M^2}\bar{\Psi}(p_1)\Gamma_\mu
(\hat{\tilde{P}}-\hat{p}_2+m)I^d\Gamma_\tau^d\Psi_c(-p_2)\,.
\eqno{(3.14)}$$
In the photon-deuteron center-of-mass frame one has:
$$\tilde{s}=\frac
14\left(\frac{s+M^2}{M}-\sqrt{\frac{s-4m^2}{s}}\frac{s-M^2}
{M}z\right)^2\,,
z=\cos(\vec{q},\vec{p}_1)\,,$$
$$\tilde{P}=\sqrt{\frac{\tilde{s}}{s}}\left(\frac{s+M^2}{2M}\,,
\vec{n}\frac{s-M^2}{2M}\right)\,.
\eqno{(3.15)}$$

Let us compare the expressions for the pole diagram provided by
the Feynman amplitude and the dispersion relation one.
 In the deuteron rest frame
the denominator of the Feynman amplitude (3.2) is:
$$
2(m^2-p^2)=2(2Mp_{20}-M^2)=2M(2p_{20}-M).\eqno{(3.16)}
$$
Corresponding denominator related to the pole term in the dispersion
relation amplitude is equal to:
$$
\tilde s-M^2=4p^2_{20}-M^2=(2p_{20}+M)(2p_{20}-M).\eqno{(3.17)}
$$
Let us stress that the momentum $p$ is different in
both techniques, namely, $p=P-p_2$
for the Feynman integral and $p=\tilde P-p_2$ for the dispersive one
and, correspondingly,
the  functions $\phi,\varphi,  f_1$ and $f_2$ are different too.
However, in the non-relativistic limit, $(2p_{20}+M) \simeq 2M$,
the dispersive and Feynman pole
amplitudes coincide with each other.

Since the intermediate particles in the dispersive technique are
mass-on-shell, the functions   $\phi$ and $\varphi$ in (3.5) depend on
 $\tilde P ^2$ only  $(
 \tilde P^2=\tilde{s})$. Their numerical values are given
 in \cite{akms}.
Form factors  $f_1^N$ and $f_2^N$ depend on $q^2$ only.

These functions can be expressed
in a standard way in terms of    the
electric, $f_e^N$,  and magnetic, $f_m^N$, nucleon form factors:
 $$
f_1^N=f_m^N,\;\;\;
f_2^N=\frac{2m}{4m^2-q^2}(f_e^N-f_m^N)\,.
\eqno{(3.18)}$$
In equation (3.8) $q^2$ is equal to zero.

\subsection{ Nucleon-Nucleon Final State Interaction}

We calculate here  the triangle  diagram of Fig. 11b, with
nucleon--nucleon interactions in  the  final state. In
the triangle diagram  the nucleonic degrees of freedom are taken into
account only; contribution of non-nucleonic degrees of freedom is
considered in the next Section.  Although  we consider
here the one-channel nucleon-nucleon amplitude, the developed method
can be easily generalized for the case of
multichannel scattering in the final
state, such as  the scattering in $^3S_1-^3D_1$ waves.

 We classify partial $\gamma d \to pn$ amplitudes according to the
final state $pn$-waves ($pn$ system in the state $b$ is determined by
 (2.4) and (2.33)). The amplitude $\gamma d \to pn$ with the $pn$ final
state interaction is written as
$$
A_{tr}^b=\xi_\tau \epsilon_\mu M_{\tau\mu}^b\,,
\eqno{(3.19)}$$
where
$$
M_{\tau\mu}^b=\hat{g}^T({\bf 1}-B)^{-1}\Delta_{\tau\mu,\alpha'\beta'}
(\bar{\Psi}(p_1)
I_{ij}^b\tilde{Q}^b_{\alpha\beta}\Psi_c(-p_2))\,.
\eqno{(3.20)}$$
The amplitude $\Delta_{\tau\mu,\alpha \beta }$
corresponds to the triangle diagram contribution,
 Fig. 11b, while  $M_{\tau\mu}^b$
is a sum of graphs with a full set of the final state
nucleon  rescattering diagrams. Vertex matrix $\hat g^T$ and loop
diagram matrix $B$ are determined by the $pn$ scattering amplitude, see
eqs. (2.30) -- (2.34).
  $\Delta_{\tau\mu,\alpha'\beta'}$ can be written in the
 form analogous  to $\hat{g}$:
$$
\Delta_{\tau\mu,\alpha \beta }=\left(\begin{array}{c}
\Delta_{\tau\mu,\alpha \beta }^1\\
\Delta_{\tau\mu,\alpha \beta }^2\\
\vdots
\end{array}\right)\,.
\eqno{(3.21)}$$

As in the calculation
of the pole diagram, we start with the Feynman
amplitude for  $\Delta_{\tau \mu,\alpha \beta }$:
$$
\Delta_{\tau\mu,\alpha \beta }^i=\int\frac{d^4p'_1}{(2\pi)^4i}
\frac{-Sp(\Gamma_\tau^d I^d(\hat{p'}+m)
\Gamma_\mu(\hat{p'}_1+m)G_iQ^b_{\alpha \beta }I_b(-\hat{p'}_2
+m))}{(m^2-p'^2)(m^2-p'^2_1)(m^2-p'^2_2)}. \eqno{(3.22)}$$
After summing over isotopic indices, one gets
$$
Sp(I^d\Gamma_\mu I_b)=(f_1^p+(-1)^If_1^n)\gamma_\mu+
(f_2^p+(-1)^If_2^n)(p+p_2)_\mu\equiv\tilde{\Gamma}_\mu\,.
\eqno{(3.23)}$$

To make a transition from the Feynman amplitude  (3.22) to the
dispersion relation one, it is necessary to carry out the following
steps:

1. To expand the numerator of (3.22) in  external
four-vectors, for which we choose $ P$ and $P_f=p_1+p_2$:
$$-Sp\,
(\Gamma_\tau^d I^d(\hat{p'}+m)
\tilde
{\Gamma}_\mu(\hat{p'}_1+m)G_iQ^b_{\alpha\beta}I^b(-\hat{p'}_2
+m))=\sum_{j} d_j^iF_{\tau\mu,\alpha\beta}^j\,.
\eqno{(3.24)}$$
Here $F_{\tau\mu,\alpha\beta}^j$ is an expression built up
       from the vectors $P,P_f$  and tensors $g_{\tau\mu},
\varepsilon_{\mu\tau\alpha\beta}$. The procedure for constructing
invariant functions $d^i_j$ and the explicit form  of
$F_{\tau\mu,\alpha\beta}^j$ for various $pn$ scattering
channels is given in \cite{alex2}.

2. Using (3.24), to represent the expression for
$\Delta_{\tau\mu,\alpha \beta }^i$ in the form:
$$
\Delta_{\tau\mu,\alpha \beta }^i=\sum_{j}D_j^i
F_{\tau\mu,\alpha\beta}^j\,.
\eqno{(3.25)}$$
For the presentation of $D_j^i$  as dispersion relation amplitudes,
 the double discontinuity of the triangle diagram is calculated,
 i.e.  the intermediate two-particle state
is considered as a real one,
with the neutron-proton energy $\sqrt{\tilde s}$ before and $\sqrt{s'}$
after the interaction $(\tilde s,s'\geq 4m^2)$. This procedure
 corresponds to the replacement:  $$
\frac{1}{(m^2-p'^2)(m^2-p'^2_1)(m^2-p'^2_2)}\rightarrow\frac 14 (2\pi
i)^3\delta(m^2-p'^2)\theta(p'_0)
\eqno{(3.26)}$$
$$\times\delta(m^2-p'^2_1)\theta(p'_{10})\delta(m^2-p'^2_2)\theta(p'_{20})
\equiv\rho(\tilde s,s',z)\,.$$

3.To reconstruct the function $D_j^i$
using the dispersion integral with a double discontinuity
as an integrand, :
$$D_j^i=\int_{4m^2}^{\infty}\frac{d\tilde s\,\,ds'}{\pi^2}
\frac{disc\,\,disc \,D_j^i}{(\tilde s-M^2)(s'-s)}\,.
\eqno{(3.27)}$$
Here
$$disc\,\,disc D_j^i=\int\frac{d^4p_1}{(2\pi)^4}\rho(\tilde s,s',z)
   d_j^i(\tilde s,s',z)\,.
\eqno{(3.28)}$$
Using the results obtained by the derivation
of (3.14), we perform integration over $s$ in (3.27).  This gives:
$$
D_j^i=\int\frac{ds'}{\pi}\int_{-1}^{1}\frac{dz}{2}\rho(s')
\frac{2 d_j^i(\tilde{s},s',z)}{(\tilde{s}-M^2)(s'-s)}
\,.\eqno{(3.29)}$$
$\rho(s')$ is given by (2.7), $\tilde{s}$ is defined according to
(3.15),  $s=P_f^2$,
$z=\cos(\vec{q},\vec{p'}_1)$.
 Invariant functions $D_j^i$
 determine the function $\Delta_{\tau \mu,\alpha \beta}$ (3.25) and,
consequently, the amplitude $M^b_{\tau \mu}$.

\subsection{Photodisintegration Amplitude with $\Delta(1232)$ in the
Intermediate State}

In this Section we include into the photodisintegration amplitude
the $\Delta$-isobar produced in the intermediate state.
 The importance of an isobar in the deuteron photodisintegration was
emphasized in  numerous papers (see, for example,
\cite{aren2},\cite{ogawa},\cite{anas}). The isobar reveals itself in
 two ways: \\
1) As an $N\Delta$ intermediate state in the  final state
nucleon-nucleon rescatterings
 (dashed block in Fig. 1b,c);\\
2) In the prompt photoproduction, with a subsequent transition
$N\Delta \to pn$ in  the triangle diagrams of Fig. 1e-type.

Here we calculate the photodisintegration amplitude
with  $N\Delta(1232)$ in the intermediate state, taking into account
 the coupled channels $^1D_2(NN)$ $-^5S_2(N\Delta)$,
$^3P_2(NN)$ $-^5P_2(N\Delta),$ and
$^3F_3(NN)$ $-^5P_3(N\Delta)$.

1) {\bf $\Delta$-isobar, $\gamma N$ partial width and the coupling}

We start the treatment of $\Delta(1232)$ in the deuteron
photoproduction amplitude with a consideration of the
transition coupling for   $N\gamma\rightarrow\Delta$,
which  determines  the process of Fig. 1e.  The
  propagator of the stable $\Delta$-isobar, with mass $m_{\Delta}$, is
equal to $$ \frac{(\hat{p}+m_\Delta)(-g_{\alpha\beta}^\bot+\frac
13\gamma_\alpha^\bot\gamma_\beta^\bot)}{m_\Delta^2-p^2}\,.
\eqno{(3.30)}$$
Re-definition of the propagator for the case
of a decay into  nucleon and  photon consits in a replacement:
$$
(m_\Delta^2-p^2)^{-1}\rightarrow(m_\Delta^2-p^2-B_\gamma(p^2))^{-1}\,,
\eqno{(3.31)}$$
where $B_\gamma(p^2)$ is the photon--nucleon loop diagram.
 The real part of this diagram re-determines $m_\Delta$,
and the imaginary part of $B_\gamma(p^2)$  gives the partial width
$\Gamma ^w_\gamma$ for the decay
 $\Delta \to \gamma N$:
$$\frac{(\hat{p}+m_\Delta)(-g_{\alpha\beta}^\bot+\frac
13\gamma_\alpha^\bot\gamma_\beta^\bot)}{m_\Delta^2-p^2
-im_\Delta\Gamma^w_\gamma(p^2)}\,.
\eqno{(3.32)}$$
Thus, we have to calculate $B_{\gamma}(p)$. The
following expression stands for the $\Delta$-isobar propagator with
$B_{\gamma}(p)$ considered as a perturbative correction:
$$
\frac{(\hat{p}+m_\Delta)(-g_{\alpha'\beta}^\bot+\frac
13\gamma_{\alpha'}^\bot\gamma_\beta^\bot)}{m^2_\Delta-p^2}
\int\frac{d^4q}{(2\pi)^4i}\,\Gamma_{\alpha\mu}^\Delta\,\frac{(\hat{k}+
m)}{m^2-k^2}\,\Gamma_{\beta\mu'}^\Delta \,\frac{g_{\mu\mu'}}{-q^2}$$
$$\times\frac{(\hat{p}+m_\Delta)(-g_{\alpha\beta'}^\bot+\frac
13\gamma_\alpha^\bot\gamma_{\beta'}^\bot)}{m_\Delta^2-p^2}\,.
\eqno{(3.33)}$$
The vertex operator $\Gamma_{\alpha\mu}^\Delta$
has the form \cite{gourdin}:
   $$\Gamma_{\alpha\mu}^\Delta=c_\gamma \gamma_5(\hat{q}g_{\mu\alpha}
   -q_\alpha\gamma_\mu)
\sqrt{2}T_0\,.
\eqno{(3.34)}$$
Let us decompose the vectors  $k$ and $q$ over
 vectors $p$ and  $k^\bot$ ($k^\bot$ is orthogonal to $p$):
$$q=Ap+k^\bot\,,\qquad k=Bp-k^\bot\,,
\eqno{(3.35)}$$
   where
$$A=\frac{p^2-m_\Delta^2}{2p^2}\,,\qquad
B=\frac{p^2+m_\Delta^2}{2p^2}\,.$$
To calculate the imaginary part of (3.33) the following
replacement should be made:

$$\frac{1}{(-q^2)(m^2-k^2)}\rightarrow\frac{1}{2}(2\pi
i)^2\delta(m^2-k^2)\theta(k_0)\delta(-q^2)
\theta(q_0)\,.
\eqno{(3.36)}$$
When integrating over  $k^\bot$, the terms
proportional to $k^\bot$ and
$k^{\bot 3}$ vanish, while the  quadratic terms should be replaced as
  follows:
 $$
k_\alpha^\bot k_\beta^\bot\rightarrow\frac
13\left[-\frac{(p^2- m^2)}{4p^2}\right]g_{\alpha\beta}^\bot=\frac 13
k_\bot^2g_{\alpha\beta}^\bot \,.
\eqno{(3.37)}$$
Using the  equality
$$\left(-g_{\alpha\beta}^\bot+\frac 13
\gamma_\alpha^\bot\gamma_\beta^\bot\right)
\left(-g_{\beta\varepsilon}^\bot+
\frac 13 \gamma_\beta^\bot\gamma_\varepsilon^\bot\right)=
\left(-g_{\alpha\varepsilon}^\bot+\frac 13
\gamma_\alpha^\bot\gamma_\varepsilon^\bot\right)\,.$$
one obtains:
$$
\frac{(\hat{p}+m_\Delta)\left(-g_{\alpha'\beta'}^\bot+\frac 13
\gamma_{\alpha'}^\bot\gamma_{\beta'}^\bot\right)}{m_\Delta^2-p^2}
\hat{p}\left(A^2B p^2-\frac 23 Ak_\bot^2-\frac 13 Bk_\bot^2\right)
$$
$$\times\, i\frac{p^2-m^2}{16\pi m_\Delta
p^{1/2}}\, c_\gamma^2\,.
\eqno{(3.38)}$$
Fixing $p^2 = m_\Delta^2$ in the numerator, we have for
the width  $\Gamma^w_\gamma$:
$$
\Gamma^w_\gamma=\frac{1}{8\pi}\left(\frac{m_\Delta^2-m^2}
{2m_\Delta}\right)^3\frac{2m_\Delta^2+\frac 23 m^2}{m_\Delta^2}
\, c_\gamma^2\,.
\eqno{(3.39)}$$
After presenting $c_\gamma$ as $c_\gamma=(e/m_\pi)C$,
we get  $C=0.30$ (at $\Gamma^w_\gamma/\Gamma^w_{tot}=0.006$
and $\Gamma^w_{tot}=115$ MeV), in accordance with
 refs. \cite{pdg} and \cite{ogawa}.

2) {\bf $pn$-production in the waves $^1D_2(NN)$, $^3P_2(NN)$ and
$^3F_3(NN)$}

The amplitude of the process $\gamma d \to pn$ is given by eq. (3.20).
For the waves $^1D_2(NN)$, $^3P_2(NN)$ and $^3F_3(NN)$),
the structure of the triangle diagram
 $\hat{\Delta}_{\tau\mu,\alpha\beta}$ is the same as
 $\hat{g}$ (which is given by (2.48)), namely:
$$
\hat{\Delta}=\left[\begin{array}{cc} \Delta^N&0\\ \Delta^t&0\\
0&\tilde{\Delta}^r\\
0&\tilde{\Delta}^\Delta\end{array}\right]\,.
\eqno{(3.40)}$$

We have two types of triangle diagrams: with nucleons in the
intermediate state, $\Delta ^N$ and $\Delta ^t$ (Fig. 11b), and
with $N\Delta$ intermediate state, $\tilde \Delta ^r$ and
$\tilde \Delta ^\Delta$  (Fig. 11c).
The  calculation of $\Delta^i$ is given in Section 3.2.
The scheme of calculation of $\tilde{\Delta}^i$ is presented below. The
Feynman integral corresponding to the diagram of Fig. 11c is:\\
$$\tilde{\Delta}_{\tau\mu,\alpha\beta}^i=
\int\frac{d^4p'_1}{(2\pi)^4i}
\frac{-Sp(\Gamma_\tau^d\tilde{I}^d(\hat{p'}+m)\Gamma_{\mu\varepsilon}
^\Delta(\hat{p'}_1+m_\Delta)\tilde{P}_{\varepsilon\varepsilon'}
G_iQ^a_{\alpha\beta,\varepsilon'}T_0(-\hat{p'}_2+m))}
{(m^2-p'^2)(m_\Delta^2-p'^2_1)(m^2-p'^2_2)}\,,
\eqno{(3.41)}$$
where $\Gamma_{\mu\varepsilon}^\Delta$ is the vertex for the $\Delta$-
photoproduction (3.34).  The spin-dependent numerator of the isobar
propagator $\tilde{P}_{\varepsilon\varepsilon'}$ is:
$$
\tilde{P}_{\varepsilon\varepsilon'}=-g_{\varepsilon\varepsilon'}
^\bot+\frac 13 \gamma_\varepsilon^\bot\gamma_{\varepsilon'}^\bot
\,,
\eqno{(3.42)}$$
where
$$g_{\varepsilon\varepsilon'}^\bot=g_{\varepsilon
\varepsilon'}-\frac{p'_{1\varepsilon}
p'_{1\varepsilon'}}{p'^2_1}\,,\qquad \gamma_\varepsilon^\bot=
\gamma_\varepsilon-\frac{\hat{p'}_1p'_{1\varepsilon}}{p'^2_1}\,.$$

The expression (3.41) is written for a stable isobar. In the case
 of an unstable isobar decaying into pion
  and nucleon, the pole term  should be replaced as:
$$
(m_\Delta^2-p'^2_1)^{-1}\rightarrow(m_\Delta^2-p'^2_1-B_\pi(p'^2_1))^{-1}\,,
$$
where the function $B_\pi(p'^2_1)$ is determined by the
pion--nucleon loop diagram. The real part of this diagram
re-determines the mass, and the imaginary part gives the width
$\Gamma^w_\pi$ of the
decay  $\Delta\rightarrow N\pi$:
$$Im\,B_\pi(p'^2_1)=m_\Delta\Gamma^w_\pi(p'^2_1)\,.$$
Using the Lehman representation, the isobar propagator can be
written as follows:
$$
(m_\Delta^2-p'^2_1-im_\Delta\Gamma^w_\pi(p'^2_1))^{-1}=\int_{(m+m_\pi)^2}
^{\infty}\frac{d\tilde{m}^2R(\tilde{m}^2)}{\tilde{m}^2-p'^2_1-i0}
\,,$$
where the weight function, $R(\tilde{m}^2)$, has the form:
$$
R(\tilde{m}^2)=\frac{1}{\pi}\frac{m_\Delta\Gamma^w_\pi(\tilde{m}^2)}
{(m_\Delta^2-\tilde{m}^2)^2+(m_\Delta\Gamma^w_\pi(\tilde{m}^2))^2}\,.
\eqno{(3.43)}$$

Below we accept   $\Gamma^w_\pi=\Gamma^w_{tot}$, and the
parametrization of ref. \cite{aren2} is used for the
momentum-dependent  width:
$$
\Gamma^w_\pi (\tilde m^2)=\Gamma^w_{tot}\frac{\tilde X^3/(1+\tilde X^2)}
{X^3_\Delta/(1+X_\Delta^2)}\,,$$
Here $\tilde X=\tilde q r/m_\pi$, $X_\Delta=q_\Delta r/m_\pi$ with
r=0.081, $\tilde q$ and $q_\Delta$ are  pion momenta in the
isobar center-of-mass  frame for the isobar masses $\tilde m$ and
$m_\Delta$, correspondingly.

Transition from the Feynman amplitude (3.41) to the dispersion
relation one makes it
 necessary to follow the prescription given in
Section 3.2. Then we get for $D^i_j$:
$$
D_j^i=\int_{(m+m_\pi)^2}^{\infty}d\tilde{m}^2R(\tilde{m}^2)
\int_{(m+\tilde{m})^2}^{\infty}\frac{ds'}{\pi}\int_{-1}^{1}\frac{dz}{2}
\frac{2d_j^i}{(\tilde{s}-M^2)(s'-s_f)}\times$$
$$\times\left(\frac{(s'-(\tilde{m}+m)^2)(s'-(\tilde{m}-m)^2)}
{s'^2}\right)^{1/2}\frac{1}{16\pi}\,.
\eqno{(3.44)}$$

\subsection{ Deuteron Photodisintegration and
Inelasticities in the Final State Waves $^1S_0$,
$^3P_0$ and $^3P_1$ }

In Section 2.7-2.8 the description of the $NN$ scattering in the waves
$^1S_0,$ $^3P_0$ and $^3P_1$ has been done, and three different
models for the mechanism of inelasticity  being suggested: production
of $\Delta(1232)$, $N^*(1440)$, or  $(\pi N)_S$-pair.
It is shown in Appendix A that the transition amplitude $NN\rightarrow
NN$ does not depend on the model used for the inelasticity, i.e. on
the type of the transition amplitude $NN\rightarrow NR$,
$NR\rightarrow NN$ and $NR\rightarrow NR$, where $R=\Delta$, $N^*$, or
$(\pi N)_S$.

This means that the process of Fig. 1b  does not
depend on the  inelasticity type as well, and it may be calculated
within the technique presented in Section 2. However, the processes
of Fig.1 c,e,f, with a prompt photoproduction of a resonance or pion --
nucleon pair, may depend on the inelasticity type.
For the calculation of processes of Fig. 1c,e,f,
 it is necessary to know the resonance photoproduction amplitudes.
 It should be noted that $N^*(1400)$
has an about 50$\%$ branching rate into the channel $N\pi \pi$, so the
triangle diagram with  $N^*(1400)$ in the intermediate state
effectively describes the process of Fig. 1d.

1) {\bf Triangle diagram with $N\Delta$ intermediate state}

The vertex $\gamma N \to \Delta$ has been presented in Section 3.3.
The calculation of triangle diagram with $N\Delta$ intermediate state
is carried out in the same way as for the waves considered in previous
Section.

2) {\bf Triangle diagram with $NN^*(1400)$ intermediate state}

 The resonance $N^*(1440)$ has the same quantum numbers as a
 nucleon, $J^p=\frac 12 ^+$,  and the vertex function $F_{\gamma}$ for
the transition $N\gamma\rightarrow N^*$ is equal to:
$$ \Gamma_\mu^{N^*}=c_\gamma\gamma_\mu\tau_3\,,
\eqno{(3.45)}$$
with the following
 relation between $c_\gamma$ and the width $\Gamma_\gamma^w$of the
decay $N^* \to \gamma N$:
$$\Gamma_\gamma^w
M_{N^*}=c_\gamma^2\frac{2(M_{N^*}-m)^2}{16\pi M_{N^*}^2}
(M_{N^*}^2-m^2)\,.
\eqno{(3.46)}$$
The correlation between $\Gamma_\gamma^w$
  and $c_\gamma$ may be found in the
same way as it has been done in previous Section for $\Delta(1232)$.
   To define $c_\gamma$,  we rely upon the fact that
  the total width of $N^*$ is equal to $\Gamma_{tot}=200$ MeV. Let
  the ratio $\Gamma^w_\gamma/\Gamma^w_{tot}$ be the same for  $p\gamma$
  and $n\gamma$ channels, being equal to 0.10\% \cite{pdg}. Using
(3.46), we get $c_\gamma=0.3$.  Experimental data \cite{pdg} on  $N^*$
  do not allow to define the coupling with a high accuracy.

3) {\bf Triangle diagram with $N(\pi N)_S$ intermediate state}

Following the model presented in Section 2.7, we
approximate a pion-nucleon pair by the quasi-resonance $R$. Then the
vertex function for the transition $\gamma N\rightarrow R$ can be
written in the form:
$$\Gamma^R_{\mu}=c^R_\gamma\gamma_\mu\gamma_5\tau_3\,.
\eqno{(3.47)}$$
The coupling  $c^R_\gamma$ is found by calculating
the $\gamma p\rightarrow\pi^+ n$ cross section:
  $$\sigma(\gamma
p\rightarrow\pi^+n)=\int_{-1}^{1}dz \frac{1}{4J}
\frac1{16\pi s_{\gamma p}}[(s_{\gamma
p}-(m_{\pi}^2+m^2))^2-4m^2m_\pi^2]^{1/2}$$
$$\times\frac{c_\pi^2c_\gamma^2}{(M_R^2-s_{\gamma p})^2+M_R^2
\Gamma_R^{w\,2}}
g_{\mu\mu'}^{\bot\bot}Sp\{(\hat{k}+m)\gamma_\mu\gamma_5(\hat{k}+
\hat{k}_{\pi}+M_R)\,
\eqno{(3.48)}$$
$$\times(\hat{k}+m)\gamma_{\mu'}\gamma_5(\hat{k}+\hat k_{\pi}+M_R)\}
\frac 23 \,,$$
where $q$ and $k$ are photon and neutron momenta, $s_{\gamma p}$ is the
$\gamma p$ energy squared, and
$$J=2(s-m^2)\,,\quad g_{\mu\mu'}^{\bot\bot}=g_{\mu\mu'}-
\frac{k_\mu k_{\mu'}}{k^2}-\frac{q_\mu^\bot q_{\mu'}^\bot}
{q^{\bot 2}}\,,\quad q_\mu^\bot=q_\mu-k_\mu\frac{(qk)}{m^2}\,.$$
The parameters $c_\pi$, $M_R$ and $\Gamma_R$ standing for the
quasi-resonance  are introduced in Section 2.7. According to
 \cite{arm}, the $\gamma p\rightarrow\pi^+n$ cross section
has a resonance pick related to the $\Delta$-isobar and a smooth
background. We relate this background to the quasi resonance
production of the $(\pi N)_S$-pair.
The quasi-resonance width $\Gamma_R^w$ is calculated with the simplest
ansatz about the vertex of the decay    $R\rightarrow\pi N$:
  $$\Gamma_\pi=c_\pi\vec{\tau}\,.
\eqno{(3.49)}$$
The  fit provides
the following parameters
 $c_\gamma=0.5$, $c_\pi=0.7$, $M_R=1.85$ GeV.

For all  considered waves ($^1S_0$, $^3P_0$ and $^3P_1$) and for
all types of the models, the result is the following: processes with
inelastic production in the triangle-graph intermediate state provide
small contribution, in other words, the contribution of diagrams 1c, 1e
and 1f is not large, while the main  one comes from the
diagram of Fig. 1b. For example, the maximum value of the triangle
graph contribution in the wave $^1S_0$ comes at $E_\gamma=30-50$ MeV,
and it does not exceed 5\% of $\sigma(^1S_0)$ (see below).

 \section{Results and Discussion}

\subsection{Polarization Averaging and  Cross Section Calculations}

Within the used normalization, the differential unpolarized cross
section has the form:
$$
d\sigma=\frac{| {A}|^2}{J} d\Phi\,,      \label{3.1}
\eqno{(4.1)}
$$
where $ {A}$ is the whole disintegration amplitude, which is
built up of the amplitude $A=A^{pole}+\sum _b A^b_{tr}$ by averaging
over the spins of incident particles and summed over spins of final
particles. The summation $\sum _b$ is performed over all considered
states  $^1S_0$,
$^3S_1-^3D_1$, $^3P_0$, $^1P_1$, $^3P_1$, $^3P_2$, $^1D_2$, $^3D_2$ and
$^3F_3$.

The two-particle phase space factor, $d\Phi$, and the flux factor
$J$ are defined as:
$$
d\Phi=\frac{d^4p_1d^4p^2}{(2\pi)^6}(2\pi)^4\delta^4(P-p_1-p_2)
\delta(m^2-p_1^2)\delta(m^2-p_2^2)\,,         \label{3.2}
\eqno{(4.2)}
$$
  $$J=4((Pq)^2-m^4)^{1/2}\,.$$
For non-polarized deuteron we need to make the replacement:
$$\xi_\tau\xi_{\tau'}\rightarrow -\frac
 13 g_{\tau\tau'}^\bot=-\frac13\left(g_{\tau\tau'}-\frac{P_\tau
P_{\tau'}}{P^2}\right)\,.\eqno{(4.3)}$$
For the non-polarized photon, it is necessary to average over the two
polarization states:
$$\varepsilon_\mu\varepsilon_{\mu'}\rightarrow
-\frac{1}{2} g_{\mu\mu'}^{\bot\bot} =
-\frac
      12\left(g_{\mu\mu'}-\frac{P_\mu P_{\mu'}}{P^2}-
\frac{q_\mu^\bot q_{\mu'}^\bot}{q^{\bot 2}}\right)\,,\eqno{(4.4)}$$
where
$q^\bot=q-P(Pq)/P^2$, and $q_\mu g_{\mu\mu'}^{\bot\bot}=0$ and
$g_{\mu\nu}^{\bot\bot} g_{\nu\xi}^{\bot\bot}=g_{\mu\xi}^{\bot\bot}$.
 It means that the combination
$g_{\mu\mu'}^{\bot\bot}A_{\mu'}$ which is used for the calculation of
the cross section is gauge invariant.

\subsection{Photodisintegration Cross Section:
Nucleonic Degrees of Freedom}

At the first stage of our investigations \cite{alex2}, we have
calculated the amplitude $\gamma d \to pn$ taking into account
nucleonic degrees of freedom only: the photodisintagration amplitude is
described by the sum of diagrams 1a and 1b and parameters of $G$
functions are restored on phases of $NN$ scattering only.

 Fig. 12a presents the calculation result obtained in such
approximation: solid line stands for the pole diagram contribution
(Fig. 1a), while dashed line describes the cross section
$\sigma (\gamma d \to pn)$ calculated with both diagrams 1a and 1b.

One can see that for nucleonic degrees of freedom the
photodisintegration reaction is described at $E_\gamma \le 10$ MeV
only, that agrees with the result of ref. [37].

One should note that the diagram of final state interaction does not
change the result obtained in the framework of the impulse
approximation. This is a consequence of the completness condition for
two sets of wave functions: plain waves and those with the $pn$
interaction. The proximity of solid and dashed lines is a criterion of
self-consistency of our calculations. It should be noted that in the
region of very small $E_\gamma$ the situation changes: the diagrams of
Fig. 1a and 1b cancel each other, due to the orthogonality of the
deuteron wave function and the wave function of continuous spectrum.

\subsection{Contribution of Inelastic Channels}

Inelastic processes are important in the waves  $^3P_2$, $^1D_2$
and $^3F_3$ (the production of $\Delta$-isobar and in the waves
$^1S_0$, $^3P_0$ and $^3P_1$ (the production of $N^*(1400)$ or
$(\pi N)_S$-state).

The calculation result for $\sigma(\gamma d \to pn)$, with the  account
of inelasticities in the waves $^3P_2$, $^1D_2$ and $^3F_3$, is shown
in Fig. 12b (dashed-dotted line). At $E_\gamma \simeq 300$ MeV the
calculated cross section has a bump, but it is lower that the
experimental one. At  $10 \le E_\gamma \le 100$ MeV the dashed-dotted
line is located below the curve which is given by the pole diagram,
that is due to the destructive interference. It should be noted that in
these waves the main contribution comes from the diagram of the prompt
$\Delta$-isobar production (Fig. 1e), i.e. from the process $\gamma N
\to N\Delta$.

Inelasticities in the waves $^1S_0$, $^3P_0$ and $^3P_1$ lead to the
considerable increase of the cross section $\sigma(\gamma d \to pn)$
$10 \le E_\gamma \le 200$ MeV: because of that, the cross section
$\sigma_{calc}(\gamma d \to pn)$ satisfactorily describes  the
experimental data up to $E_\gamma \sim 60$ MeV (solid curve in Fig.
12b).

Fig. 13 demonstrates the calculation results and the
experimental data [10-13] for differential cross sections at
$E_\gamma=20,60,90$ MeV. An agreement is seen at $E_\gamma=20,60$ MeV.
However in the region $E_\gamma >60$ MeV there is no agreement with
experimental data.
One may believe that this discrepancy is due to meson currents which
are not completely taken into account in our calculation.

Let us discuss in more detail the structure of processes which are
taken and are not taken into account in the approach
under discussion. In terms of the analytic structure of the
amplitude $\gamma d \to pn$, the diagrams with rescatterings do take
into account the $(s,t)$ and $(s,u)$ anomalous (Landau) singularities
(remind that $t=(q-p_1)^2$ and $u=(q-p_2)^2$). Figs. 14a,b give an
example of diagrams with such singularities:
 Fig. 14a demonstrates the diagram with the Landau
singularity in $(s,t)$-sector, while the diagram of Fig. 14b has the
Landau singularity in $(u,t)$-sector. Both these diagrams have also
singularities of triangle diagrams, which are shown in Fig. 14c,d ---
these diagrams come to being due to the shrinkage of the upper
nucleon line into a  point. The diagrams of Fig.14c-type have
anomalous singularity of the logarithm type at $t \simeq
m^2+2\mu^2+4\mu\sqrt{m\epsilon} \equiv t_{tr}$, that is rather close to
the physical region of the reaction ($\epsilon $ is deuteron binding
energy).  Similarly, the singularity at $u=t_{tr}$ is inherent to the
diagram of Fig.  14d. With final state interaction accounted for in our
approach, we also take these singularities into consideration, but not
completely, for just the same diagrams are present in the $(u,t)$-box
diagrams --- see Fig. 14e,f. These latter are in fact the meson
exchange current diagrams which are not included into our approach, and
one may believe that the lack in the cross section description at
$E_\gamma >50$ MeV should be filled in by the account of these
diagrams, together with the subsequent final state interaction (Fig.
14g).

The role of meson exchange currents in the deuteron
photodisintegration has been estimated by Laget \cite{Laget}.  Fig. 15
demonstrates the differnce $\Delta \sigma= \sigma_{exp} (\gamma d \to
pn)-\sigma_{calc}(\gamma d \to pn)$ obtained in our calculations and
the evaluation of meson exchange current contribution given in ref.
\cite{Laget} for the processes shown in Fig. 14e,f. One can see that
the estimation given in ref.  \cite{Laget} provides a reasonable
description for $\Delta \sigma$ in the region $E_\gamma=50-200$ MeV but
cannot explain the data at $E_\gamma \sim 300$ MeV. This is
exactly the region where the bump in the cross section corresponds to
the $\Delta$ isobar in the intermediate state (see Fig. 1c).
Therefore, it is natural
 to suggest that in the considered case we meet analogous
situation: the $\Delta$ isobar is produced
 in the process induced by meson exchange currents --- see Fig. 14h.
The next transition $\Delta N \to NN$ (Fig.14i) would provide a
contribution affecting the enhancement of  the cross section around
$E_\gamma \sim 300$ MeV. We believe that this is the only reasonable
explanation of the discrepancy observed in this
region: the bump in the cross section is  undoubtedly due to the
$N\Delta$ channel, and the only one which is not accounted for in our
approach is the contribution of the $(u,t)$ singularity into the
amplitude $\gamma d \to N\Delta$, i.e. the contribution of meson
currents into the process $\gamma d \to N\Delta$.

\section{Conclusion}

We have carried out the dispersion relation analysis of the nucleon --
nucleon scattering and deuteron photodisintegration amplitudes. As a
result, all the waves which are not small in the region $T_{lab}<1$
GeV are written within the despersion N/D representation.
$N$-functions which are the analogues of the potential are restored
for the waves $^1S_0$, $^3S_1-^3D_1$, $^3P_0$, $^1P_1$, $^3P_1$,
$^3P_2$, $^1D_2$, $^3D_2$ and $^3F_3$ (the $N$-functions for the
deuteron channel $^3S_1$ and $^3D_1$ were found previously in
\cite{akms}). Besides, the N/D representation is written for inelastic
amplitudes $NN \to N\Delta$ (the waves $^1D_2(NN)-^5S_2(N\Delta)$,
$^3P_2(NN)-^5P_2(N\Delta)$ and $^3F_3(NN)-^5P_3(N\Delta)$) and $NN \to
N(N\pi)_{S-wave}$ (the waves $^1S_0$, $^3P_0$ and $^3P_1$), thus
giving a complete dispersion relation representation  of the low and
intermediate energy amplitude at $T<1$ GeV.

The amplitude of the photodisintegration $\gamma d \to pn$ has been
analysed in the framework of the dispersion relation technique for
$E_\gamma <400$ MeV. The spectator mechanism of the photodisintegration
related to the interaction of photon with  one of the deuteron nucleons
is considered, the inelasticity related to the production of $\Delta$,
$N^*$ or $(\pi N)_S$ pair being included. Furthermore, using the
constructed nucleon-nucleon amplitudes, the final state interaction is
properly taken into account within the dispersion relation technique.
All these calculations are parameter-free, the necessary
characteristics of the nucleon-nucleon amplitude are directly taken
from from the experiment, that is, $\gamma N$ amplitudes, deuteron
vertices found in \cite{akms} from the fit of the deuteron form factors
at $Q^2\le 1.5$ (GeV/c)$^2$ and the $NN$ amplitudes found in this
paper. Using the language of the dispersion relation representation,
meson exchange currents are the contributions of the $(u,t)$-singular
part of the amplitude $A(\gamma d \to pn)$. The calculations
performed allow one to find out the meson exchange current
contribution: the difference between the observed
and calculated cross sections is a direct measure of it.
 The analysis shows that the spectator
mechanism for the photon-deuteron interaction, with a consistent
account of the final state interaction, adequately describes the data
at $E_\gamma < 50$ MeV. However, at $E_\gamma > 50$ MeV the meson
exchange currents begin to dominate,  providing more than a half
of the cross section at $E_\gamma \ge 100$ MeV. Comparing the results
of our calculations with those of Laget \cite{Laget}, one may conclude
that the stardard mechanism for the meson  current contribution,
namely, interaction of
incident photon with a meson which provides an exchange
 between deuteron nucleons, may explain the value of the
deuteron photodisintegration cross section at $E_\gamma <150$ MeV only,
but not around 300 MeV.
It favours the idea that the bump in $\sigma_{tot}(\gamma d\to pn)$
at $E\gamma \simeq 300$ MeV with the width about 300 MeV is due to
contribution of the $(u,t)$-singulare part of the amplitude $\gamma d
\to N\Delta$: that is a meson exchange current contributing into $NN
\to N\Delta$ when the photon interacts with $t$- or
$u$-channel meson.

At the same time, our analysis
 gives rise to the problem of meson exchange currents in the
deuteron channel, the $NN$ waves $^3S_1$ and $^3D_1$: the matter is
that deuteron form factors $A(Q^2)$ and $B(Q^2)$ are well described in
a broad interval, $0<Q^2 \le 1.5$ (GeV/c)$^2$, without an exchange
current \cite{akms}. It is possible that in this respect the deuteron
channel is a singular one, and meson exchange currents
are not large there. It is a well-known fact that in the deuteron
channel the inelasticities are strongly suppressed: we cannot
exclude that in the deuteron channel these characteristic features are
somehow related to each other. However this problem needs especial
investigation that is beyond the scope of the present study.

{\bf Acknowledgement}
We thank V.V.Anisovich, L.G.Dakhno, V.A.Nikonov and
A.V.Sarantsev for numerous discussions and useful remarks.

\section{Appendix A}

Here the $G$-function parameters are displayed, which were obtained by
fitting phases and inelasticities of the considered waves. The phase
volumes are also presented which were used in the fitting procedure.
For the deuteron waves $^3S_1$ and $^3D_1$ the parameters are given in
ref. \cite{geramb}.

The simplest way to find the $G$-functions is seen by eq. (2.22). The
$N$-functions which contain left-hand singularities related to the
$t$-channel meson exchange have the meaning of a potential of the
free-nucleon interaction. Because of that, they may change sign. This
fact has been taken into consideration by introducing two functions
$G_1$ and $G_2$, which are determined as follows:
$$
N=G_1G^1-G_2G^2.\eqno{(A.1)}
$$
In the case when inelasticity appears, the $N$-function takes a
matrix form, namely,
$$\left(\begin{array}{ll}
N_{NN\rightarrow NN} & N_{NN\rightarrow N\Delta}\\
N_{N\Delta\rightarrow NN} & N_{N\Delta\rightarrow N\Delta}
\end{array}\right),
\eqno{(A.2)}
$$
the change of sign is allowed for $N_{NN\to NN}$ only.

By fitting the phase shift data, the following circumstance have been
taken into consideration.  In eqs. (2.45) and (2.49) the $S$- and
$B$-matrices depend on the product of functions $G$ and $\rho$: $G_i
\rho_j G^k$.  For the $NN$-scattering without inelasticity, $\rho_j$
is defined by eq.  (2.7) $\rho_j=1/16\pi \,
\sqrt{(s-4m^2)/s}\equiv\rho_0 $.  But for practical use the phase
volume factor $\rho _j$ may be taken in such a way that, on the one
 hand, it would reflect the kinematics of  a chosen wave and, on the
other, it would not violate the convergence of the integral (2.49).
Let us introduce  auxilliary functions $\rho ^{(aux)}_j$ and
$G_i^{(aux)}$ which are related to $\rho _j$ and $G_i$ as follows:
$$G_i^{(aux)}\rho_j^{(aux)}G^{k{(aux)}}=G_i\rho_0G^k
\eqno{(A.3)}$$
Then $G_i=G_i^{(aux)}(\rho_j^{(aux)}/\rho_0)^{1/2}$. During the
calculations the functions $G_i^{(aux)}$ (not $G_i$) were
parametrized as follows (2.39): $G_i^{(aux)}=\sum _{n=1}^N \frac
{\gamma_n^i}{s-s^i_n}$.
The introduction of $\rho^{(aux)}$ is especially suitable for the
waves with inelasticities, when one needs to include
the inelastic threshold into the phase space.

Now consider the partial waves. For the waves $^1P_1$ and $^3D_2$ we
use the following expressions for $\rho^{(aux)}$:
$$^1P_1:\qquad\rho^{(aux)}_{^1P_1}=\frac{1}{16\pi}\frac{s-4m^2}{\sqrt{s}}
\eqno{(A.4)}$$
$$^3D_2:\qquad\rho^{(aux)}_{^3D_2}
=\frac{1}{16\pi}\frac{(s-4m^2)^2}{s^2\sqrt{s}}.
\eqno{(A.5)}$$
Parameters of $G_i^{(aux)}$ are given in Table 1.

The next step is to include resonances
$(\pi N )$ into consideration. The modified way of finding out
parameters is as follows. First, let the resonance width be zero:
$\Gamma_R^w=0$, where $R$ is one of $\Delta$, $N^*$ or $(\pi N )_S$.
Then the phase space factor, for instance for isobar $\rho_0^\Delta$,
related to the loop shown in Fig. 4b, is equal to $$
\rho_0^\Delta=\int d\Phi_{\Delta N}=\frac{1}{16\pi}\sqrt{
\frac{(s-(m_N+m_\Delta)^2)(s-(m_N-m_\Delta)^2)}{s^2}}.
\eqno{(A.6)}
$$
The trace over the loop in Fig.4b with the partial operators
$O^\kappa_{a,\alpha \beta}$ in the vertices is equal:
$$
Sp\left[O_{a,\alpha\beta}^\kappa P^{\kappa\kappa'}(-p_1+M_R)
O_{b,\alpha'\beta'}^\kappa(p_2+m_N)\right]=
\delta_{ab}T_{\alpha\beta\alpha'\beta'}\rho_a(s).
\eqno{(A.7)}
$$
Here $O^\kappa_{a,\alpha \beta}$ are partial operators which are
defined  by formulae (2.42), (2.50) and (2.53), without normalizing
factor $1/\sqrt{\Omega_a}$. $P^{\kappa \kappa'}$ is given in (3.42).

Consider now the real resonance, with $\Gamma^w_R \ne 0$, decaying into
meson and nucleon, and instead of  the diagram of Fig. 4b we use that
of Fig. 4d. There are three mass-on-shell particles in the intermediate
state, that leads to the integral over the resonance mass for the
phase volume, together with the normalization factor $\rho_a(s)$
written in (A.7). Then the phase volume for the partial state $a$
has the form:
$$
\rho_{\Delta,a}=
\int\limits_{(m_n+m_\pi)^2}\limits^{(\sqrt{s}-m_N)^2}d\tilde
s \frac{1}{16\pi}\sqrt{\frac{(s-(m_N+\sqrt{\tilde s})^2)(s-(m_N-
\sqrt{\tilde s})^2)}{s^2}}R(\tilde s)\rho_a(\tilde s),
\eqno{(A.8)}$$
and
$$
R(\tilde s)=\frac{\Gamma_R^wM_R}{(\tilde s-M_R^2)^2+
(\Gamma_R^wM_R)^2}.\eqno{(A.9)}$$

Normalizing factors in (2.42), (2.50) and (2.53) are equal
$$ \Omega_a=\rho_{\Delta,a}$$

Now let us get the equation (A.8): the three-particle phase space
factor related to the diagram of Fig. 4d. It has a form:
 $$
d\Phi_3=\frac 12\frac{d^4k_1}{(2\pi)^3}\delta\left(k_1^2-m_1^2\right)
\frac{d^4k_2}{(2\pi)^3}\delta\left(k_2^2-m_2^2\right)
\frac{d^4k_3}{(2\pi)^3}\delta\left(k_3^2-m_3^2\right)$$
$$\times(2\pi)^4\delta(P-k_1-k_2-k_3)\int
d^4k_{12}\delta^4(k_{12}-k_1-k_2)\int dm_{12}\delta\left(P-k_1-k_2
-k_3\right).\eqno{(A.10)}
$$
After multiplying it by a unity,
$$
\int d^4k_{12}\delta^4(k_{12}-k_1-k_2) \,
\int dm_{12}\delta(k_{12}-m_{12}),\eqno{(A.11)}
$$
we get a product of two two-particle phase volumes:
$$
d\Phi_3=\int\frac{dm_{12}}{\pi}\,\frac 12\,\frac{d^4k_1}{(2\pi)^3}\,
\delta\left(k_1^2-m_1^2\right)\,\frac{d^4k_2}{(2\pi)^3}\,\delta
\left(k_2^2-m_2^2\right)\,(2\pi)^4\,\delta^4(k_{12}-k_1-k_2)$$
$$\times
\frac
12\,\frac{d^4k_{12}}{(2\pi)^3}\,\delta\left(k_{12}^2-m_{12}^2\right)
\,\frac{d^4k_3}{(2\pi)^3}\,\delta\left(k_3^2-m_3^2\right)(2\pi)^4\delta^4
(P-k_3-k_4)\,.\eqno{(A.12)}
$$
They are  equal to $(m_1=m_\pi, m_2=m_N, m_3=m_N)$:
$$
\frac 12\frac{d^4k_1}{(2\pi)^3}
\delta\left(k_1^2-m_1^2\right)\frac{d^4k_2}{(2\pi)^3}\delta
\left(k_2^2-m_2^2\right)(2\pi)^4\delta^4(k_{12}-k_1-k_2)\,=$$
$$=\,\frac{1}{8\pi\sqrt{m_{12}^2}}k_{\pi N}\frac{d\Omega_{\pi
N}}{4\pi}\,=$$
$$\,=\frac{1}{8\pi\sqrt{m_{12}^2}}\sqrt{\frac{(s-(m_N+m_\pi)^2)
(s-(m_N-m_\pi)^2)}{4s}}\cdot\frac{d\Omega_{\pi N}}{4\pi},$$
$$\frac 12\frac{d^4k_{12}}{(2\pi)^3}\delta\left(k_{12}^2-m_{12}^2\right)
\frac{d^4k_3}{(2\pi)^3}\delta\left(k_3^2-m_3^2\right)(2\pi)^4\delta^4
(P-k_3-k_4)$$
$$=\,\frac{1}{8\pi\sqrt{s}}k_{N\Delta}\frac{d\Omega_{N\Delta}}
{4\pi}\,=$$
$$=\,\frac{1}{8\pi\sqrt{s}}\sqrt{\frac{(s-(m_N+m_{12})^2)(s-(m_N-m_{12})^2)}
{4s}}\cdot\frac{d\Omega_{N\Delta}}{4\pi}.\eqno{(A.13)}$$
Then, one has for the three-particle phase space factor:
$$
d\Phi_3=\int\frac{dm_{12}}{\pi}\cdot\frac{1}{8\pi\sqrt{m_{12}^2}}k_{\pi
N}\frac{d\Omega_{N\Delta}}{4\pi}
\frac{d\Omega_{\pi N}}{4\pi}.
\eqno{(A.14)}
$$
The phase space for the loop diagram of Fig. 4d is
$$
\rho_{\Delta,a}=\int d\Phi_3\frac{\alpha^2}
{(m_{12}^2-m_\Delta^2)^2+(\Gamma_\Delta^wm_\Delta)^2}
\rho_a(s,m_{12}).\eqno{(A.15)}
$$
Here  $\rho_a$ is given by (A.7). Then
$$
\rho_{\Delta,a}=
\int\limits_{(m_N+m_\pi)^2}\limits^{(\sqrt{s}-m_N)^2}
\frac{dm_{12}}{\pi}\frac{k_{\pi N}}
{8\pi\sqrt{m_{12}^2}}\frac{k_{N\Delta}}{8\pi\sqrt{s}}\frac{\alpha^2}
{(m_{12}^2-m_\Delta^2)^2+(\Gamma_\Delta^wm_\Delta)^2}
\rho_a\frac{d\Omega_{N\Delta}}{4\pi}\frac{d\Omega_{\pi N}}{4\pi}.
\eqno{(A.16)}
$$
The parameter $\alpha$ may be found, using boundary condition
$\rho_\Delta \to k_{N\Delta}/(8\pi\sqrt{s})$ at $\Gamma^w \to 0$,
providing
$$
\alpha=\frac{16\pi\Gamma_\Delta^wm_\Delta^3}{\sqrt{(m_\Delta^2-
(m_N+m_\pi)^2)(m_\Delta^2-(m_N-m_\pi)^2)}}.
\eqno{(A.17)}
$$

Coming back to eq. (A.3), one can find $G=G^{aux}(\rho_\Delta
^{aux}/\rho_\Delta)^{1/2}$ for any wave $a$. However, in this case the
triangle diagram $\tilde \Delta ^i_{\tau \mu \alpha \beta}$ (3.41)
should be calculated with non-normalized operators $O^\kappa_{a,\alpha
\beta}$, for the normalizing factor has been  already included into
$(\rho_\Delta^{aux}/\rho_\Delta)^{1/2}$.  Comparing eq.  (A.8) with
(3.43) and (3.44) for $\tilde \Delta ^i_{\tau \mu \alpha \beta}$, one
may see that the Lehman expansion takes explicitly into account the
resonance widths, when calculating the graphs 1c,e,f.

In our calculations the following parametrization of the phase space
factors were used for the waves $^1D_2$, $^3P_2$ and $^3F_3$:  $$
^1D_2:\,\rho_\Delta^{aux}=\frac{1}{16\pi}\left(0.2\frac{\sqrt{s-(2m_N+
m_\pi)^2}}{s}+20.0\frac{(s-(2m_N+0.7m_\pi)^2)^{1/4}}{s^3}\right),$$
$$\rho_N^{aux}=\frac{1}{16\pi}\frac{\sqrt{(s-4m^2)^7}}{s^3},
\eqno{(A.18)}$$

$$^3P_2:\,\rho_\Delta^{aux}=
\frac{1}{16\pi}\left(0.4\frac{\sqrt{(s-(2m_N+
m_\pi)^2)^3}}{s^2}+0.3\frac{s-(2m_N+1.8m_\pi)^2}{s^2}\right),$$
$$\rho_N^{aux}=\frac{1}{16\pi}\frac{\sqrt{(s-4m^2)^3}}{s^2},
\eqno{(A.19)}$$

$$^3F_3:\,\rho_\Delta^{aux}=\frac{1}{16\pi}\left(0.35\frac{(s-(2m_N+
m_\pi)^2)}{s^2}+0.95\frac{s-(2m_N+2m_\pi)^2}{s^2}\right),$$
$$\rho_N^{aux}(s)=\frac{1}{16\pi}\frac{\sqrt{(s-4m^2)^5}}{s^3}.
\eqno{(A.20)}$$

For the waves  $^1S_0$, $^3P_0$ and $^3P_1$ the  phase volumes
$(\rho^{aux}_R)$ are related to eq. (A.8) as
$\rho^{aux}_R(s)=\rho_\Delta(s)/s^3$, and $R(\tilde s)$ is chosen as
follows:
$$
R(\tilde s)=\sum_{i=1}^{10}\frac{\Gamma_i^wm_i}{(\tilde s-m_i^2)^2+
(\Gamma_i^wm_i)^2}, \;\;\; m_i=4.05+(i-1)0.1.
\eqno{(A.21)}
$$
The meaning of such a definition of $R$ is connected with an unclear
inelasticity nature  in the waves $^1S_0$, $^3P_0$ and $^3P_1$. When
fitting phase space data, we used as a model for inelasticity the set
of 10 resonances with masses $m_i$ and large widths $\Gamma^w_i$, these
widths defined by the decay vertices to nucleon and meson,
$c_{\pi_i}$: $c_{\pi_i}=0.028+(i-1)0.028$, and
$$
\Gamma_i^w=\frac{1}{16\pi m_i^2}c_{\pi,i}^2\left((m_i+m_N)^2-m_\pi^2
\right)\sqrt{\frac{(m_i^2-(m_N+m_\pi)^2)(m_i^2-(m_N-m_\pi)^2)}{m_i}}.
\eqno{(A.22)}
$$
The derivation of this formula is similar to that of Section 2.3.
Parameters of $G^{(aux)}$ are presented in Table 1.

In the $^1S_0$ wave there were considered several inelasticity
variants. To get vertices $G_i$ which respond to $\Delta(1232)$,
$N^*$ or $(\pi N)_S$, one should calculate correct phase space
$\rho_\Delta $, $\rho_{N^*}$ or $\rho_{(N\pi)_S}$ using the
equations (A.7) and (A.8), with the mass $M_R$ and width $\Gamma^w_R$
related to considered variant.  Then $G$-functions are calculated
using eq.(A.3).

\newpage


\newpage

\begin{figure}
\centerline{\epsfig{file=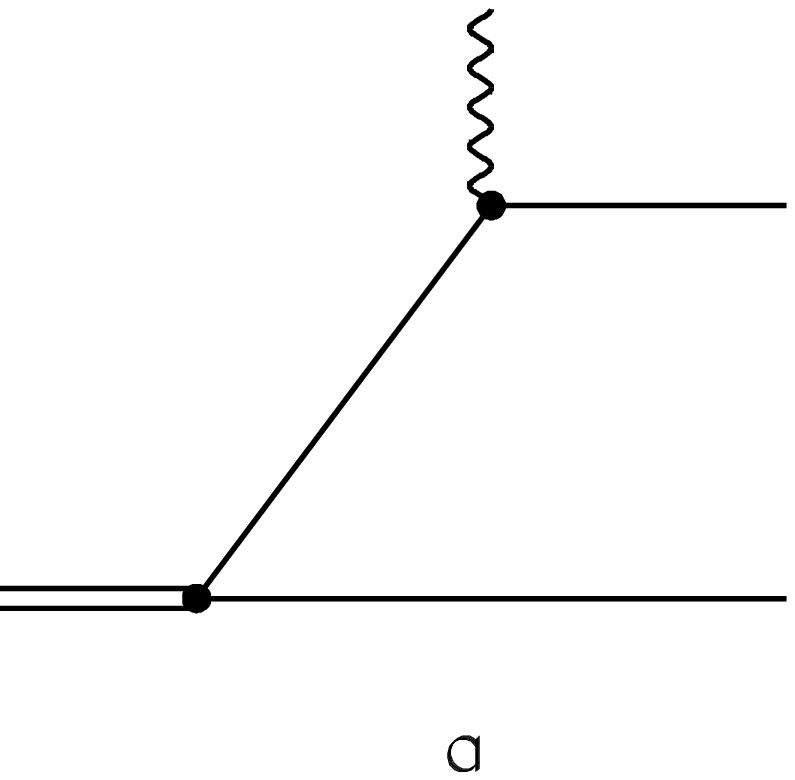,width=5cm}\hspace{2cm}
            \epsfig{file=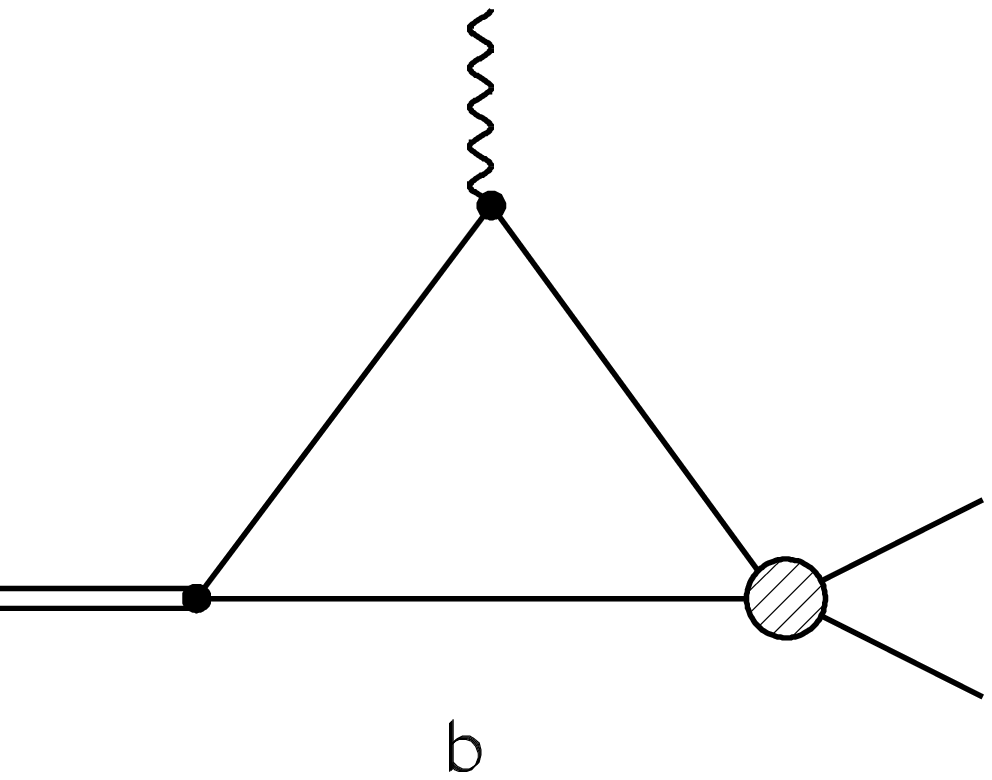,width=5cm}}
\vspace{1cm}
\centerline{\epsfig{file=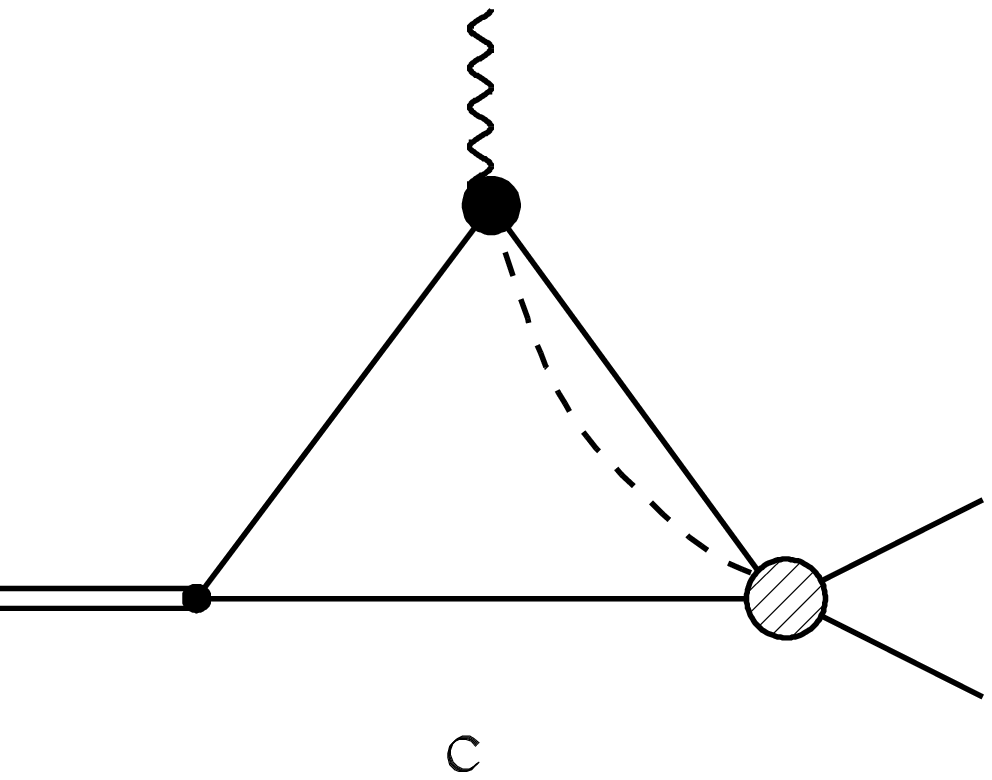,width=5cm}\hspace{2cm}
            \epsfig{file=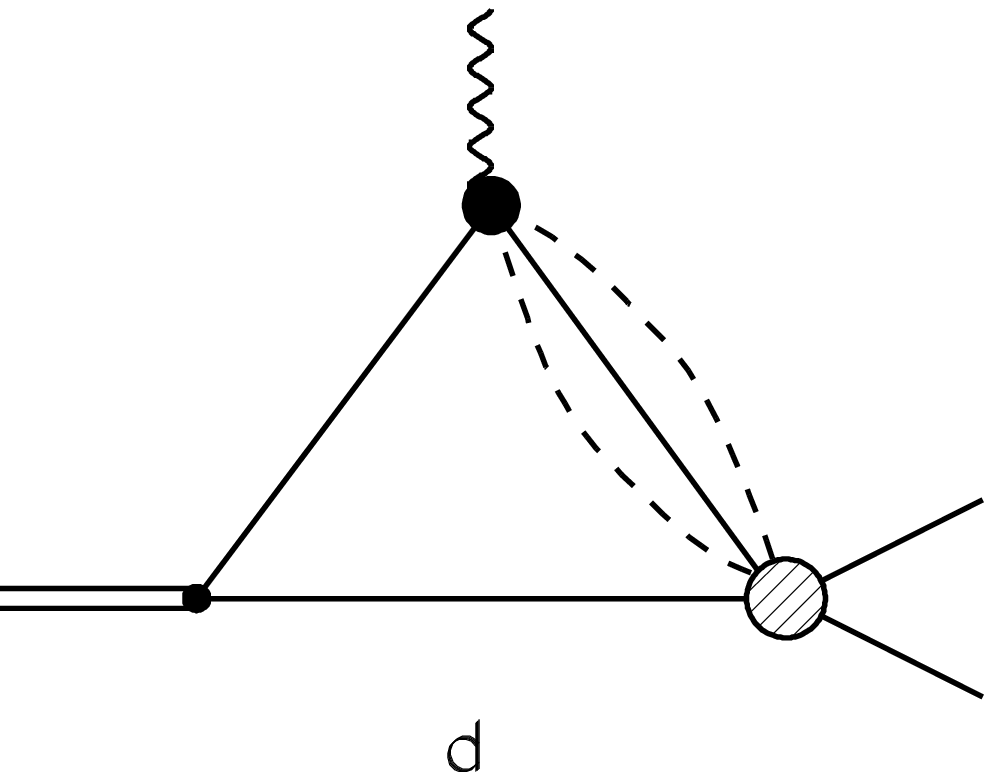,width=5cm}}
\vspace{1cm}
\centerline{\epsfig{file=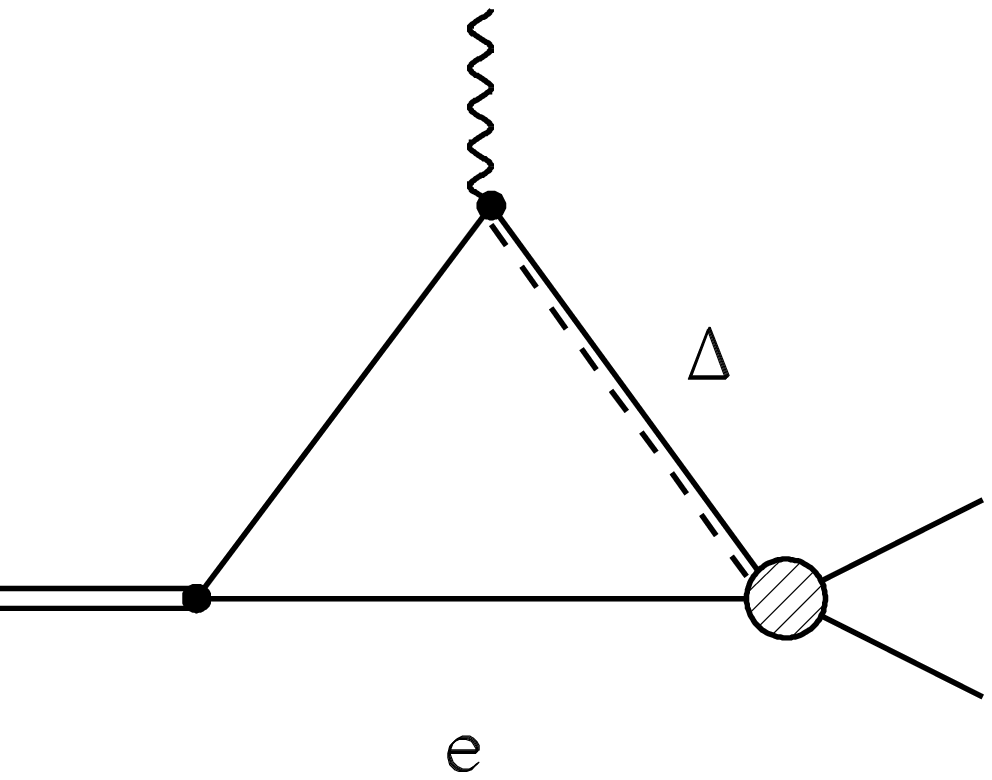,width=5cm}\hspace{2cm}
            \epsfig{file=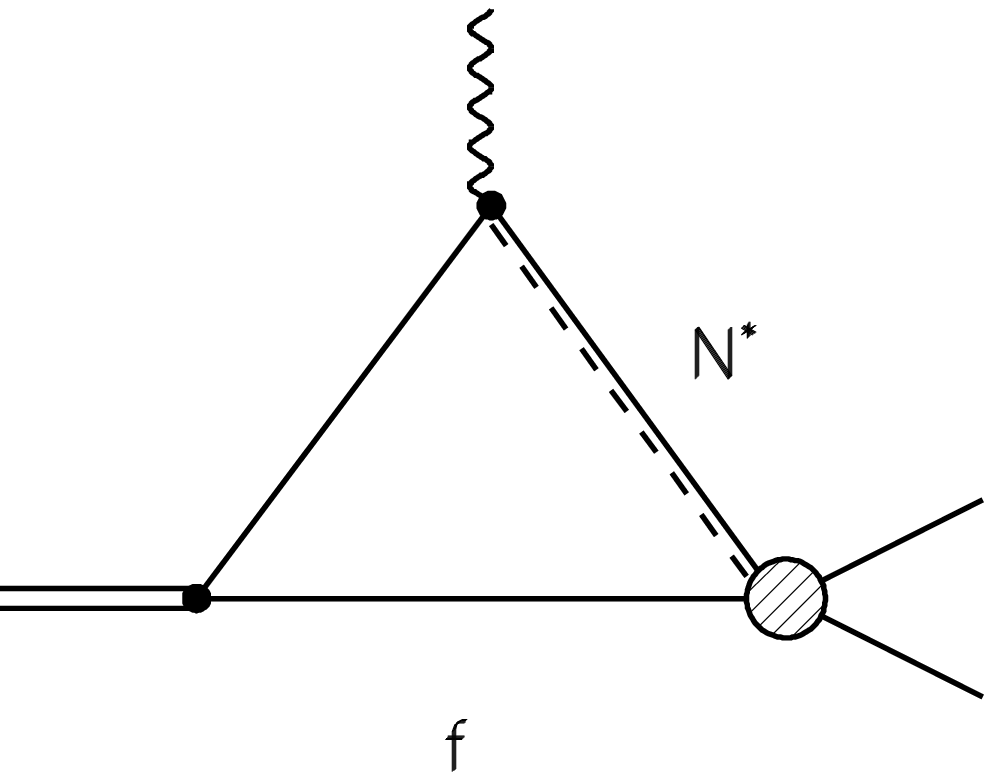,width=5cm}}
\caption{ Diagrams responsible for the deuteron disintegration
process. }
\end{figure}

\begin{figure}
\centerline{\epsfig{file=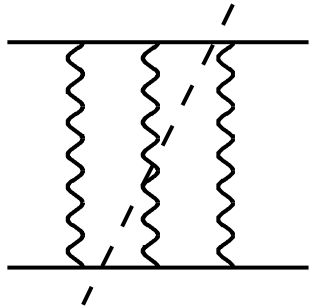,width=6cm}}
\caption{ An example of the Behte-Salpeter ladder diagram with
nucleon rescatterings:  the cutting of the diagram presents the
connected inelastic processes.}
\end{figure}

\begin{figure}
\centerline{\epsfig{file=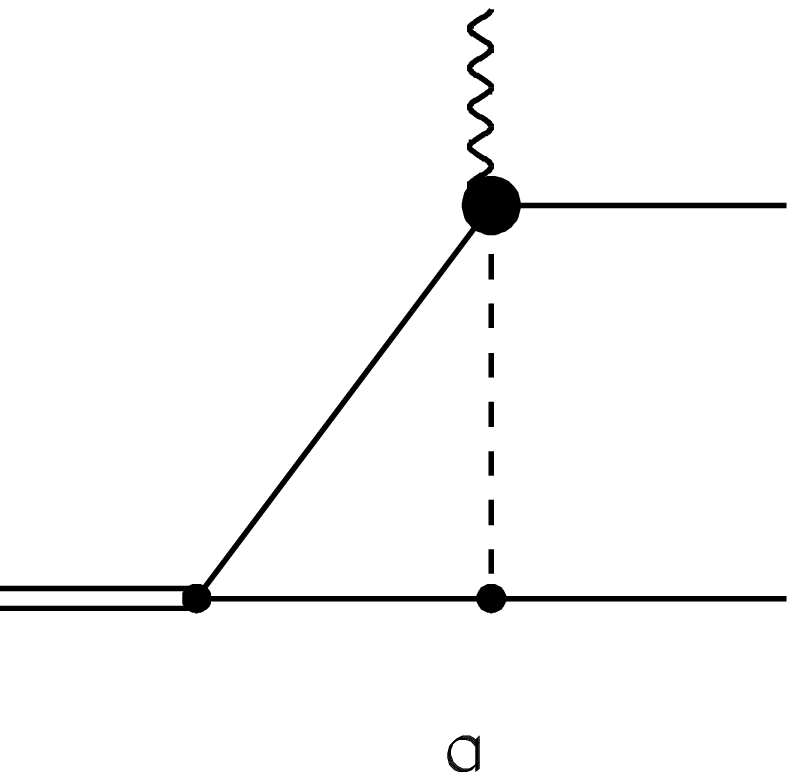,width=5cm}\hspace{2cm}
            \epsfig{file=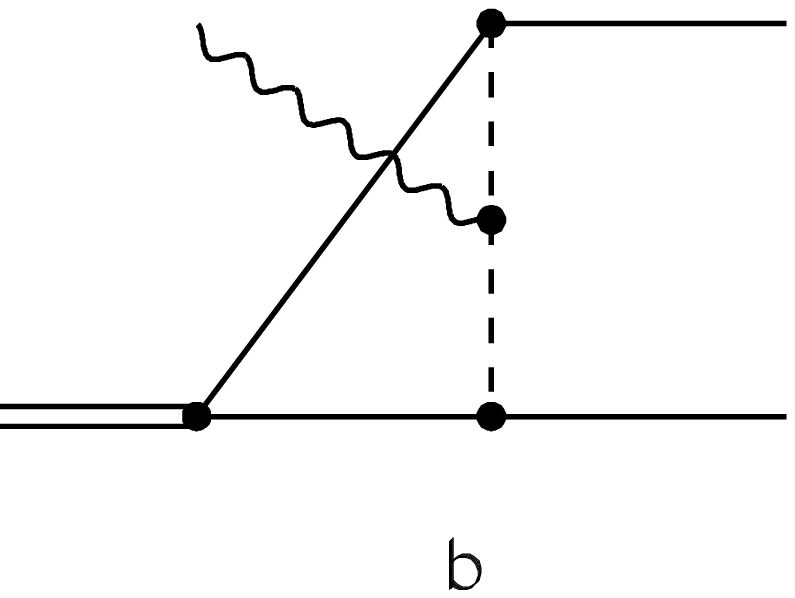,width=5cm}}
\caption{ Deuteron photodisintegration diagrams with meson exchange
currents.}
\end{figure}

\begin{figure}
\centerline{\epsfig{file=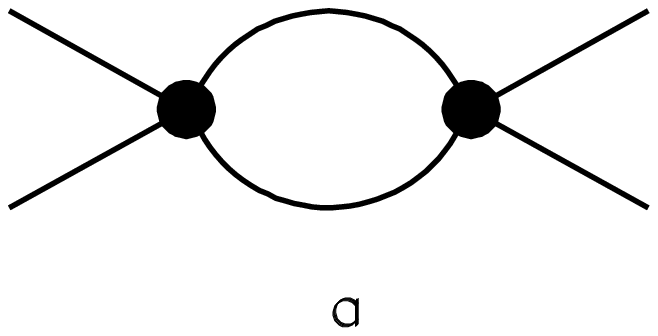,width=4cm}\hspace{0.5cm}
            \epsfig{file=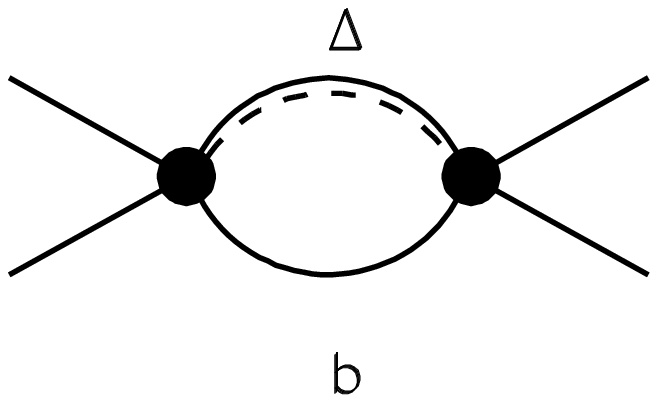,width=4cm}\hspace{0.5cm}
            \epsfig{file=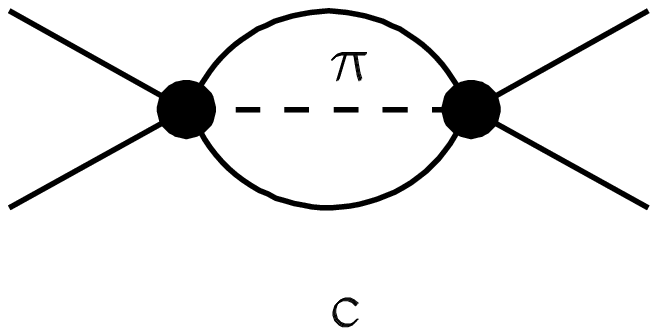,width=4cm}\hspace{0.5cm}
            \epsfig{file=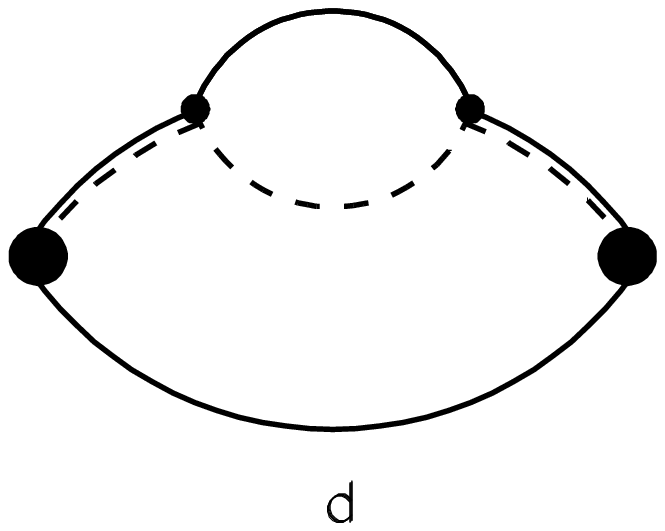,width=3cm}}
\caption{ Loop diagrams which correspond to the elastic
rescattering (a), intermediate-state production of the $\Delta$ isobar
(b), prompt pion production (c), loop diagram used in the calculation
of the two-particle phase space, with the account of the isobar width.}
\end{figure}

\begin{figure}
\centerline{\epsfig{file=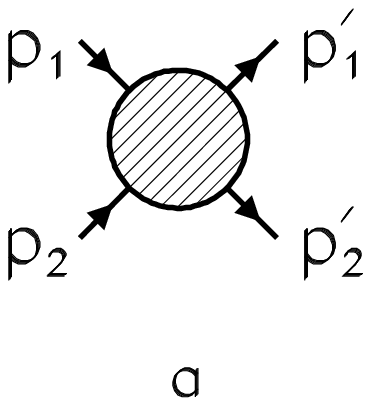,width=5cm}\hspace{2cm}
            \epsfig{file=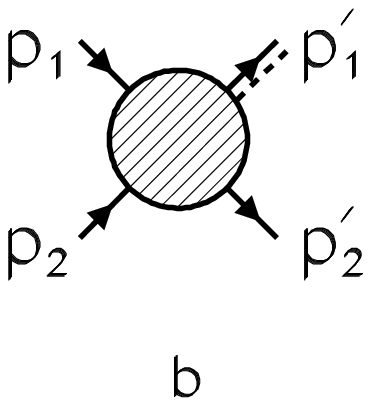,width=5cm}}
\vspace{1cm}
\centerline{\epsfig{file=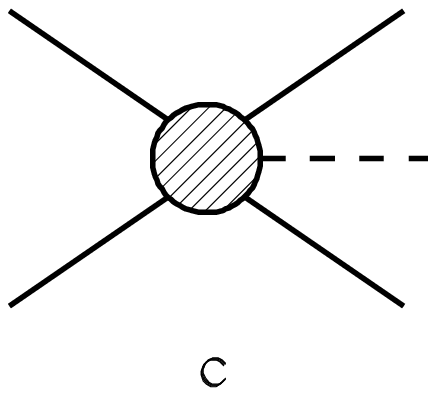,width=5cm}\hspace{2cm}
            \epsfig{file=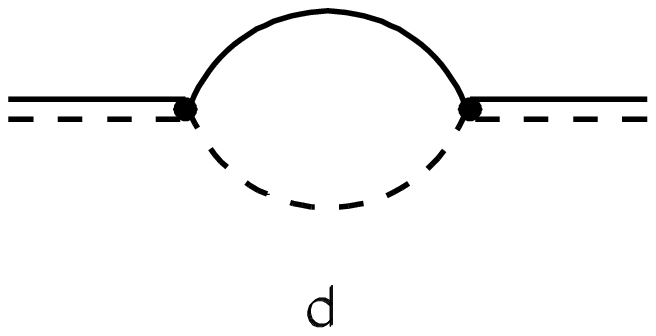,width=7cm}}
\vspace{1cm}
\centerline{\epsfig{file=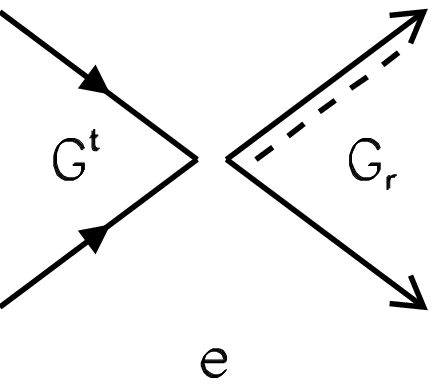,width=5cm}\hspace{2cm}
            \epsfig{file=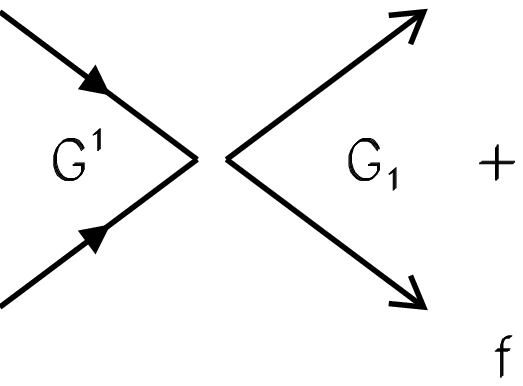,width=5.5cm}\hspace{0.5cm}
            \epsfig{file=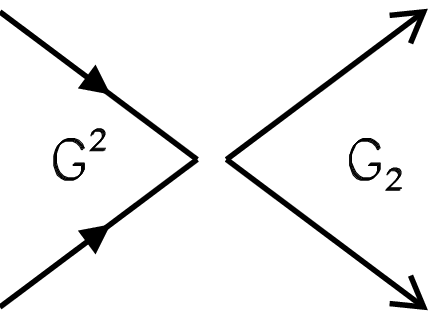,width=5cm}}
\caption{}
\end{figure}

\begin{figure}
\centerline{\epsfig{file=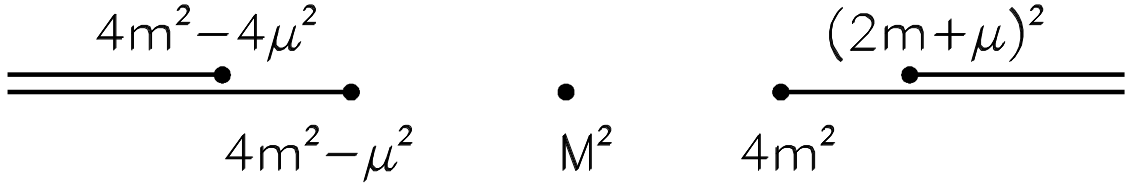,width=16cm}}
\caption{Complex $s$-plane and the position of the partial
amplitude singularities: (from right to left) right-hand side
singularities at $s=4m^2,\ (2m+m_\pi)^2\ ...$, the position of pole,
and left-hand ones at $s=4m^2-m_\pi^2,\ 4m^2-4m_\pi^2\ ...$}
\end{figure}

\begin{figure}
\centerline{\epsfig{file=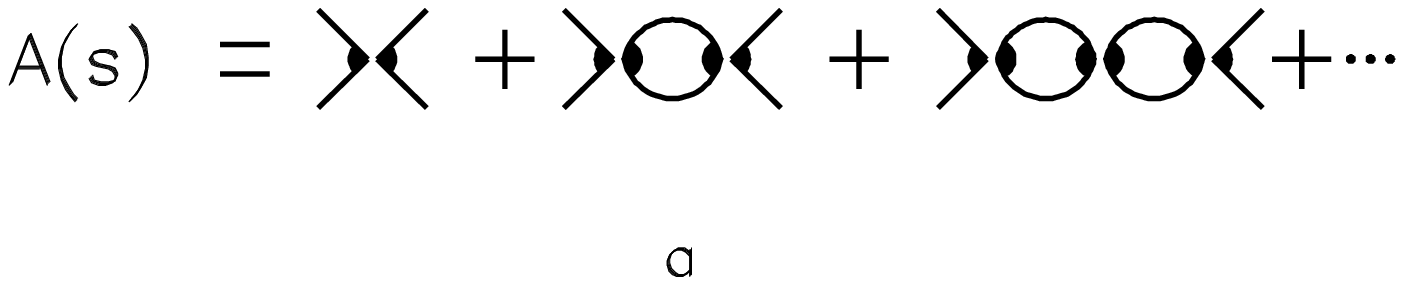,width=12cm}}
\vspace{1cm}
\centerline{\epsfig{file=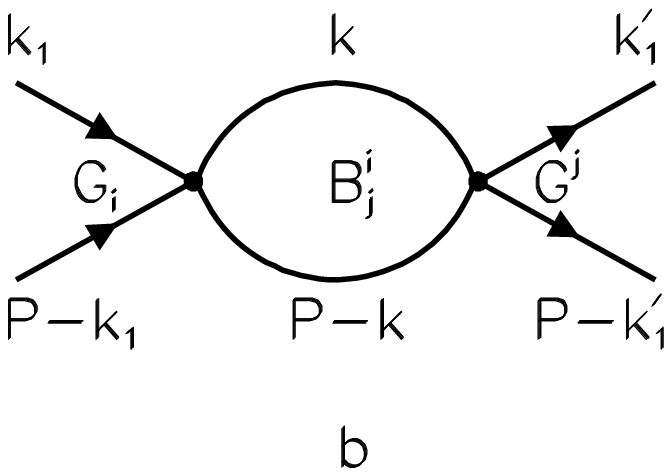,width=5cm}}
\vspace{0.5cm}
\caption{Dispersion relation loop diagrams: (a) representation of
the partial amplitude as a set of loop diagrams, (b) one-loop diagram
with right ($G_i$) and left ($G^j$) vertices: $G_jB^i_jG^i$.}
\end{figure}

\begin{figure}
\centerline{\epsfig{file=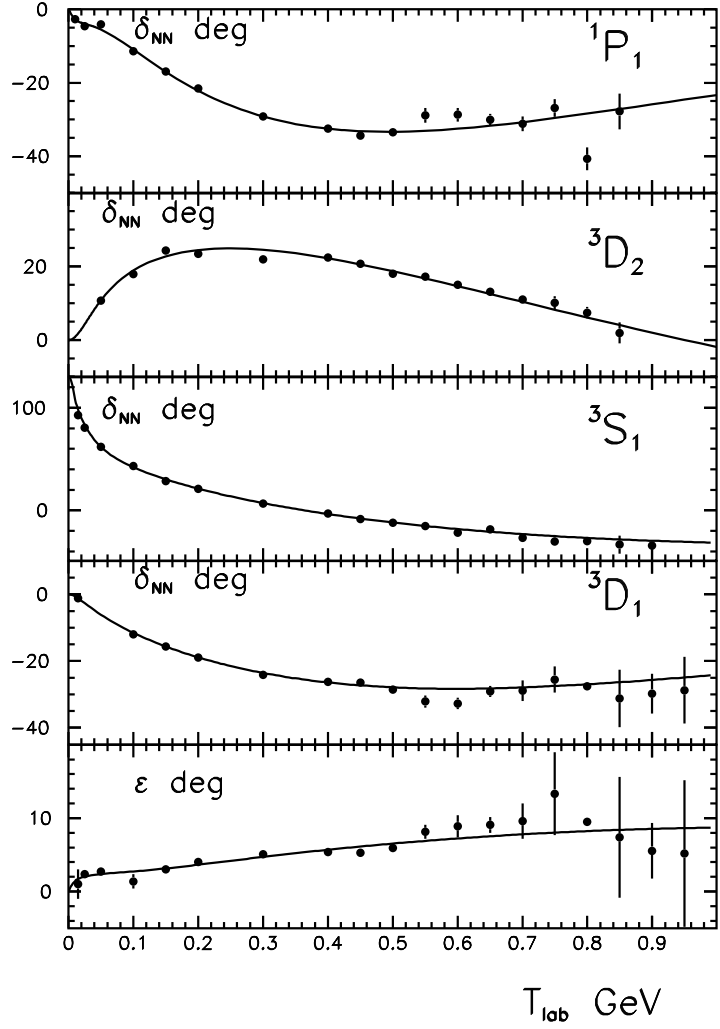,width=10cm}}
\caption{ Results of the fit for the phase shifts, $\delta_{NN}$,
of the nucleon-\-nucleon scattering. $T_{\rm lab}$ is the kinetic
energy of incident proton in the laboratory system.  The waves: {\it
a})~$^1P_1$, {\it b})~$^3D_2$; {\it c})~coupled channels
$^3S_1-^3D_1$:  phase shifts in the waves $^3S_1$ and $^3D_1$,
$\delta_{NN}$, mixing parameter $\varepsilon$.}
\end{figure}

\begin{figure}
\centerline{\epsfig{file=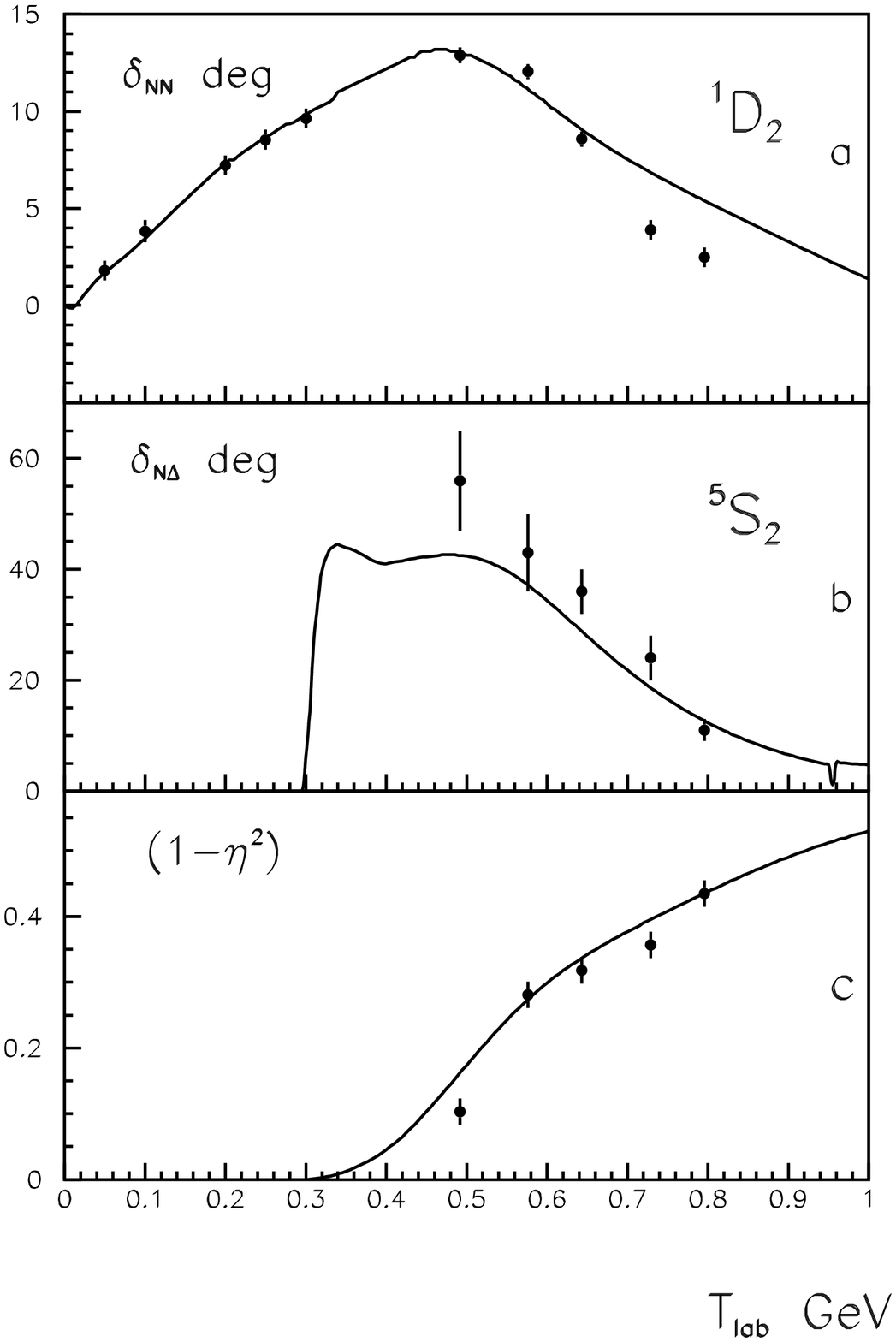,width=6.5cm}
            \epsfig{file=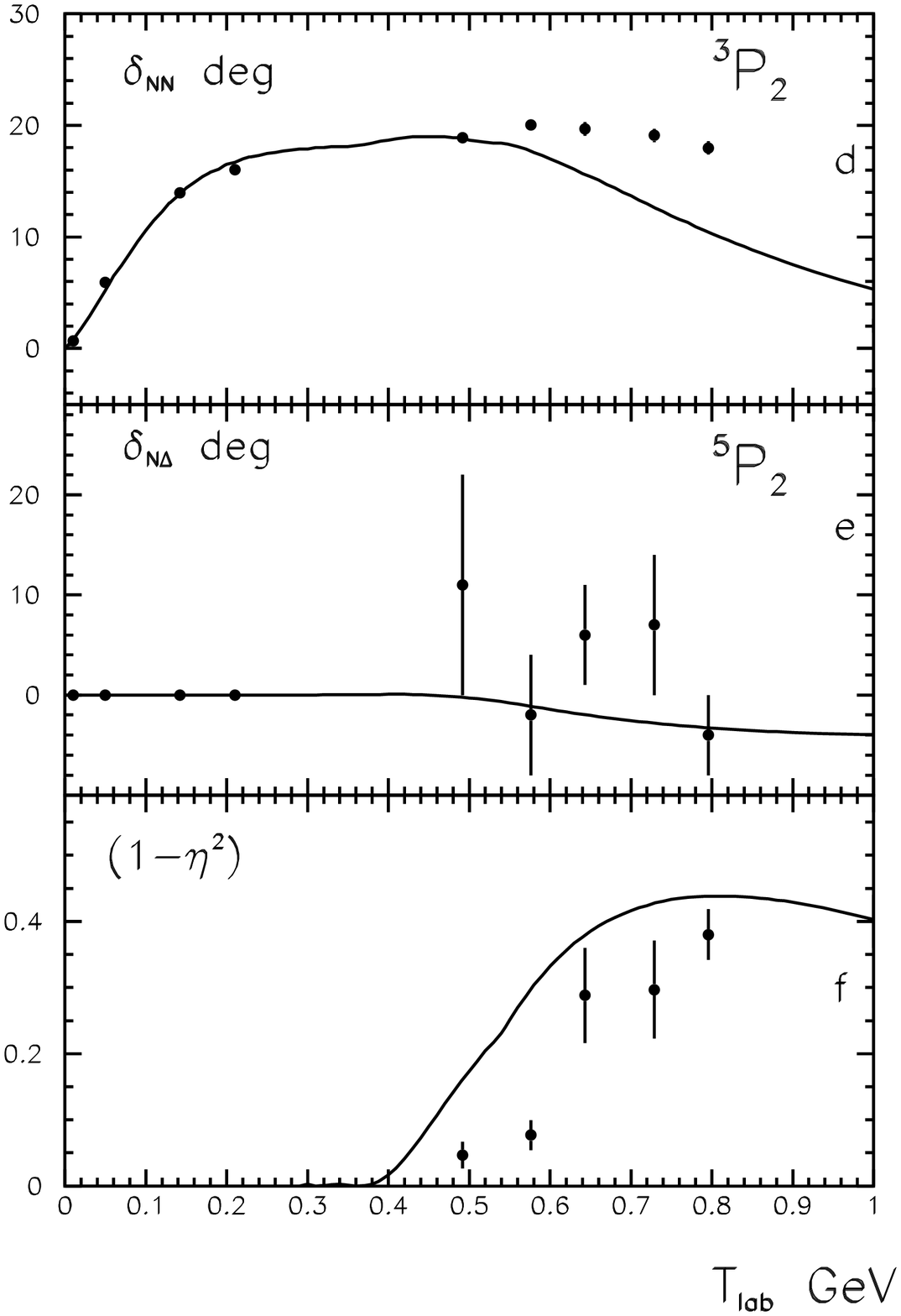,width=6.5cm}}
\vspace{0.5cm}
\centerline{\epsfig{file=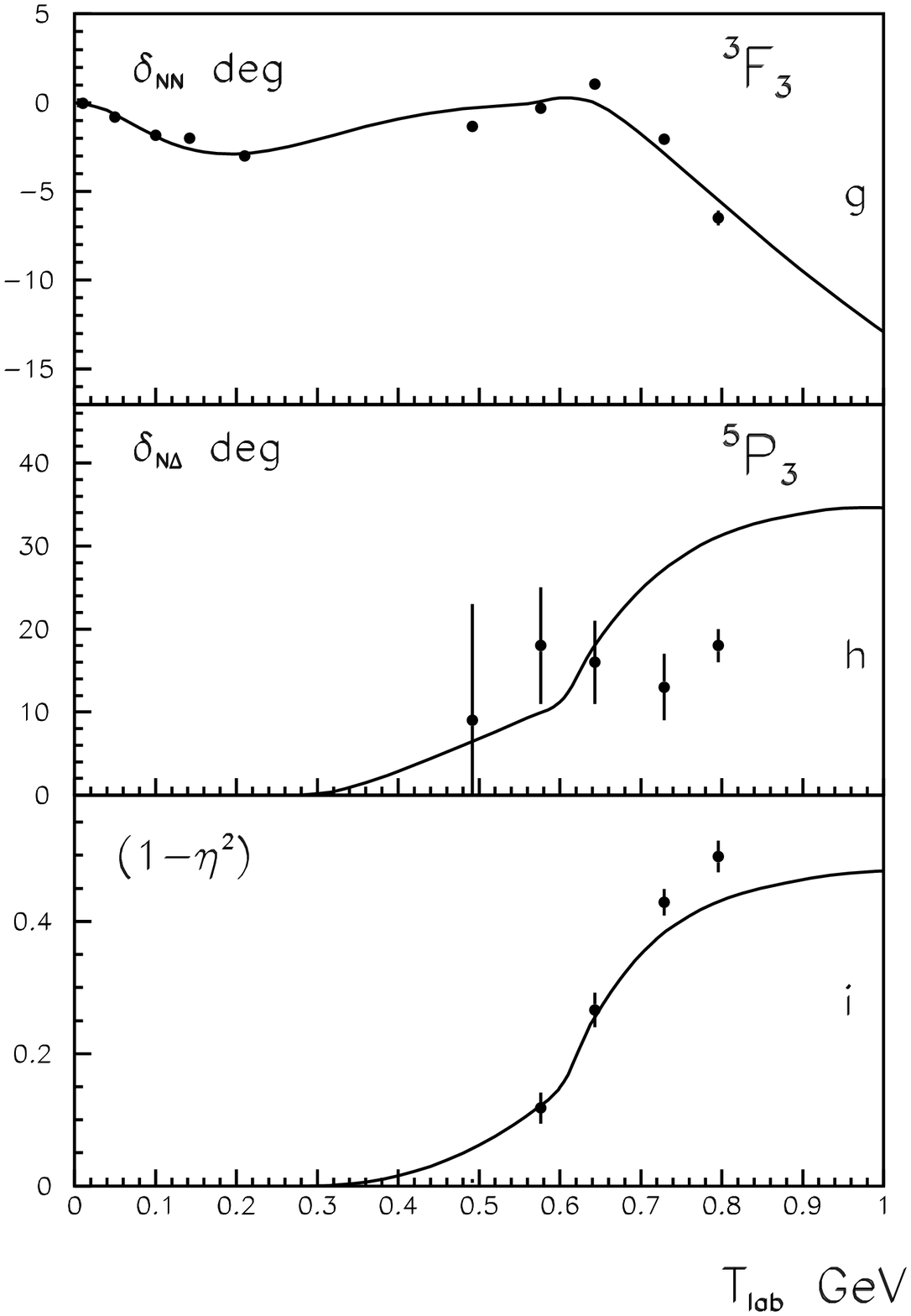,width=6.5cm}}
\caption{
Results of the fit for  phases and
inelasticities in the two-channel approach. Coupled channels
$^1D_2(NN)-^5S_2(N\Delta)$, $^3P_2(NN)-^5P_2(N\Delta)$,
$^3F_3(NN)-^5P_3(N\Delta)$:
 phase shifts in the waves:
 a)~$^1D_2(NN),\delta_{NN}$;
 b)~$^5S_2(N\Delta)$,$\delta_{N\Delta}$;
 c)~the transition amplitude squared $^1D_2(NN)-^5S_2(N\Delta)$,
$(1-\eta^2)$;
 d)~$^3P_2(NN)$,
 e)~$^5P_2(N\Delta)$,
 f)~the amplitude squared $^3P_2(NN)-^5P_2(N\Delta)$;
 g)~$^3F_3(NN)$,
 h)~$^5P_3(N\Delta)$,
 i)~$^3F_3(NN)-^5P_3(N\Delta)$.}
\end{figure}

\begin{figure}
\centerline{\epsfig{file=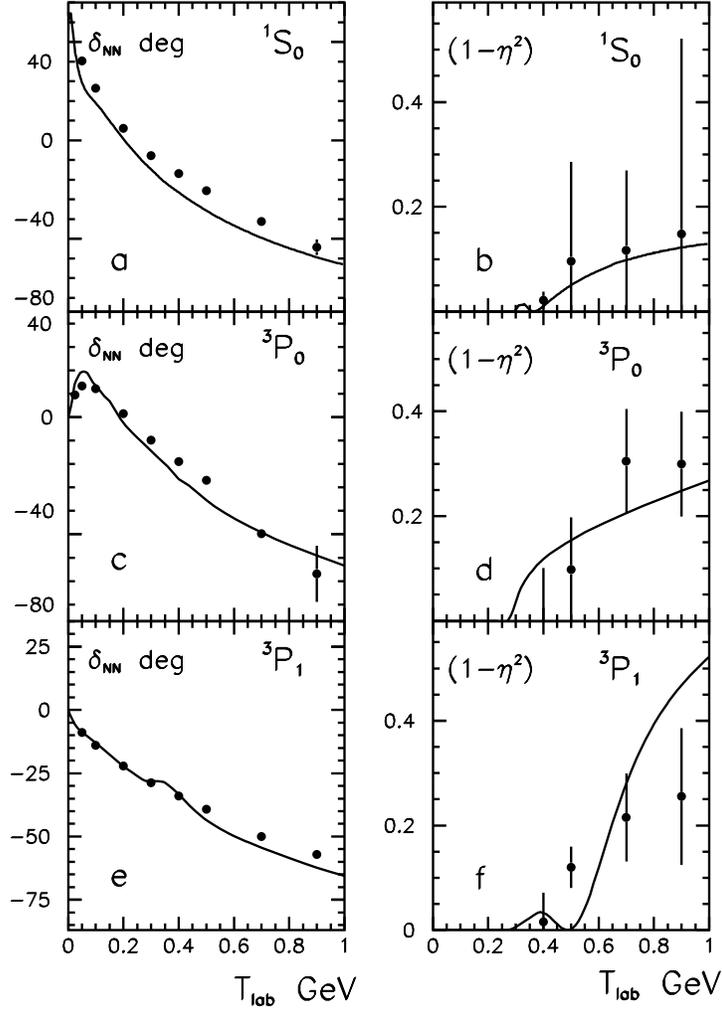,width=10cm}}
\caption{ Results of the fit for the phases  and inelasticities
in the two-\-channel approach for the waves $^1S_0$, $^3P_0$, $^3P_1$:
{\it a})~phase shifts in the wave $^1S_0$, $\delta_{NN}$, {\it b})~
inelasticity in the wave $^1S_0$, $(1-\eta^2)$; {\it c})~phase
shifts in $^3P_0$, {\it d})~inelasticity in $^3P_0$;
{\it e})~phase shifts in  $^3P_1$, f)~inelasticity in
$^3P_1$. }
\end{figure}

\begin{figure}
\centerline{\epsfig{file=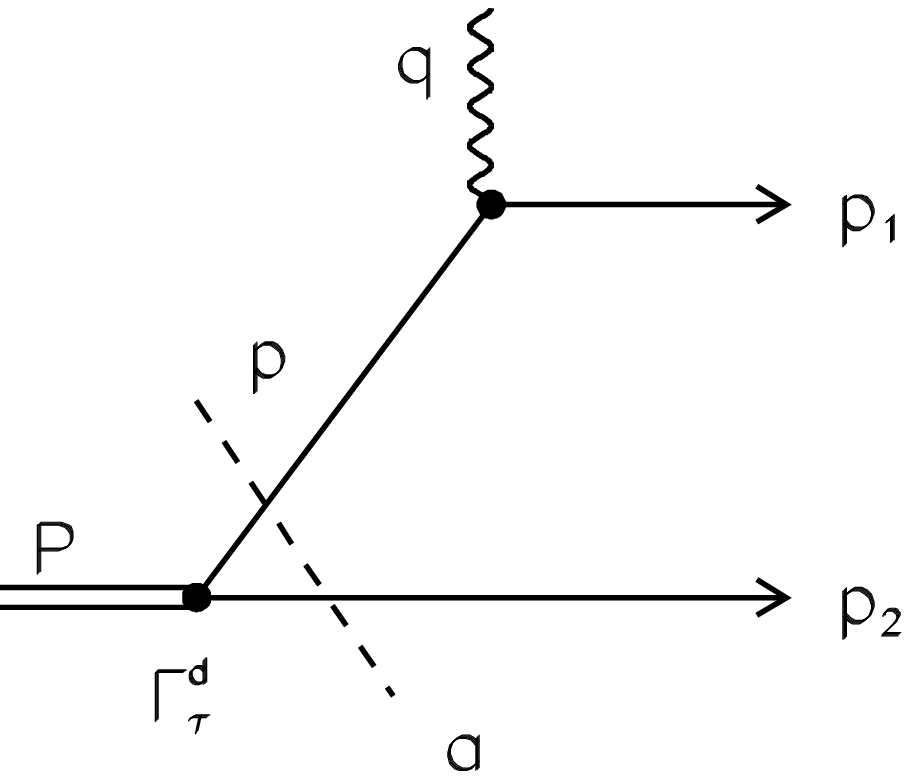,width=6cm}\hspace{1cm}
            \epsfig{file=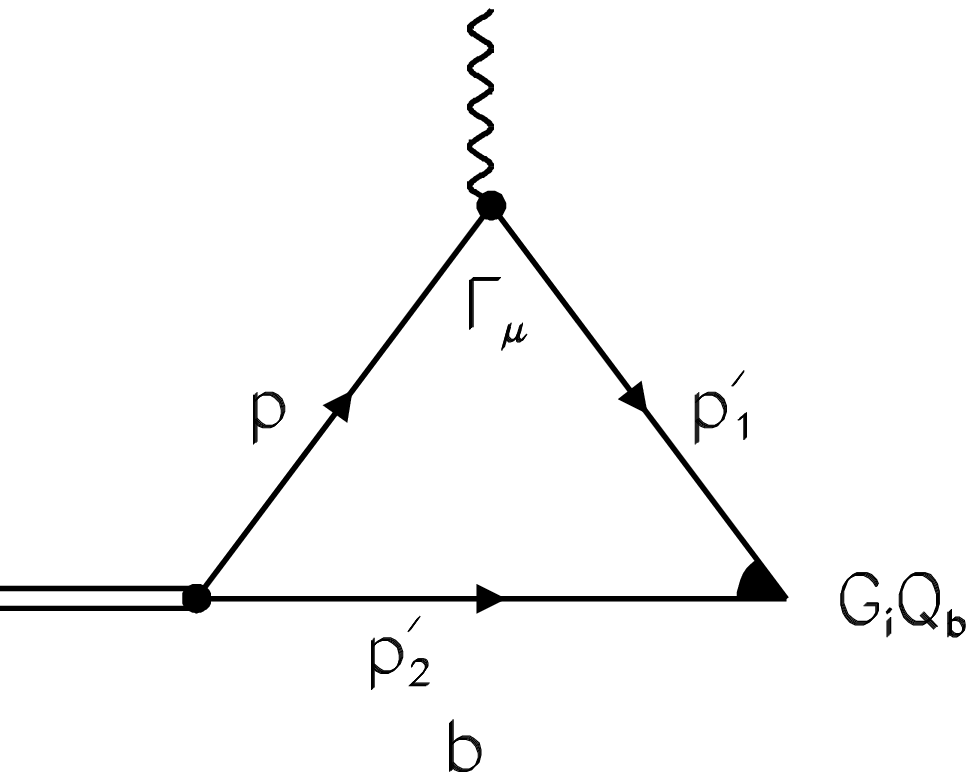,width=6cm}\hspace{1cm}
            \epsfig{file=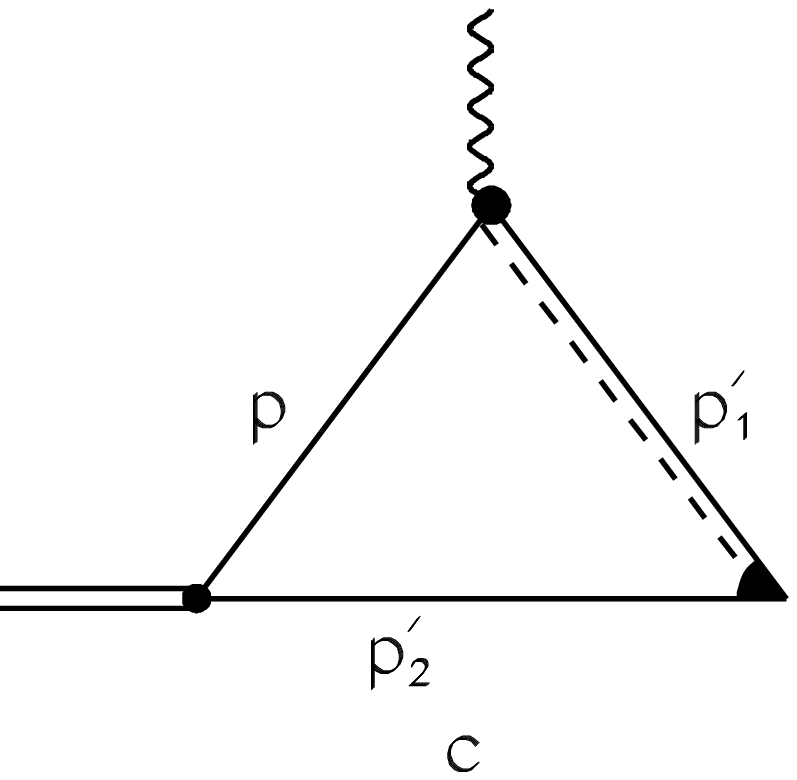,width=6cm}}
\caption{}
\end{figure}

\begin{figure}
\centerline{\epsfig{file=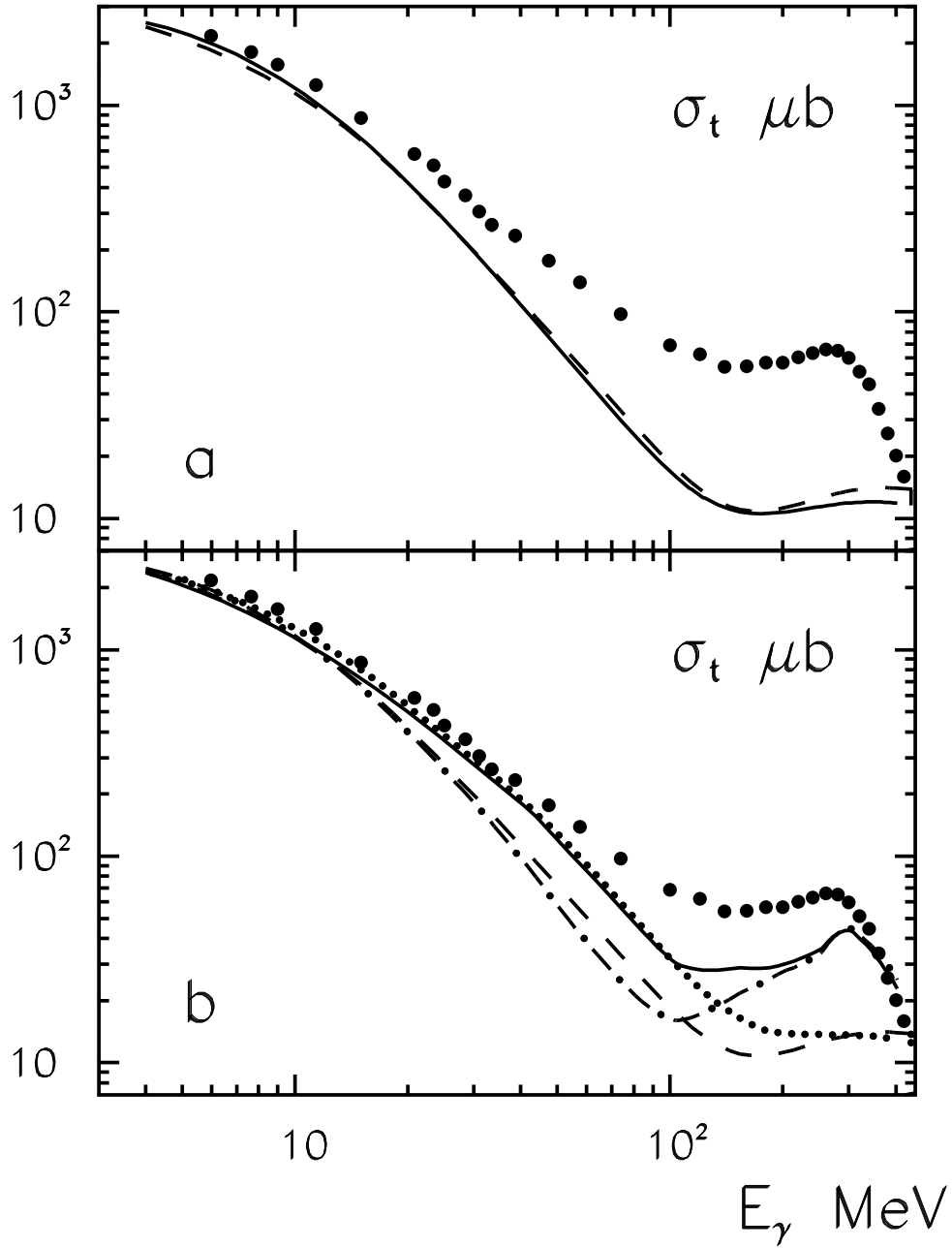,width=14cm}}
\caption{ Total cross section, $\sigma_t$, for the deuteron
photodisintegration: a)~the contribution  of the impulse approximation
diagram 1a to the $\sigma_t$ (dashed line), $\sigma_t$ with FSI
(nucleon degrees of freedom) (solid line); b)~the contribution of the
impulse approximation diagram (dashed line); $\sigma_t$ with the
inelasticity in the waves $^1D_2(NN)$, $^3P_2(NN)$, $^3F_3(NN)$
(dot-\-dashed line); $\sigma_t$ with the inelasticity in the waves
$^1S_0(NN)$, $^3P_0(NN),$ $^3P_1(NN)$ (two dot-\-dashed line); the
complete $\sigma_t$ (solid line). $E_\gamma$ is the photon energy in
laboratory system.}
\end{figure}



\begin{figure}
\centerline{\epsfig{file=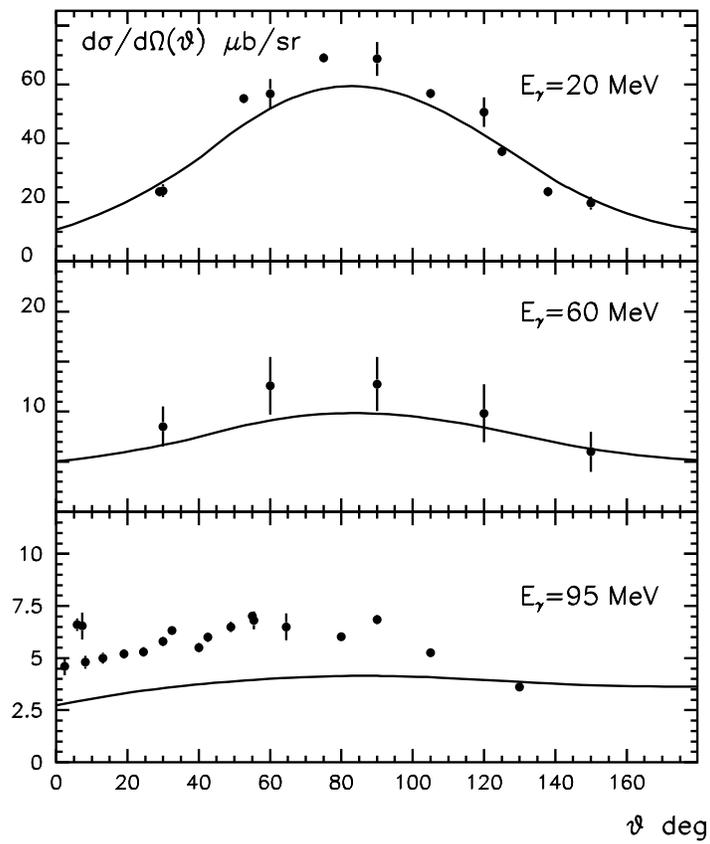,width=10cm}}
\caption{ Differential cross sections
$\frac{d\sigma}{d\Omega}(\Theta)$ at $E_\gamma=20$~MeV,
$E_\gamma=60$~MeV, $E_\gamma=95$~MeV. }
\end{figure}

\begin{figure}
\centerline{\epsfig{file=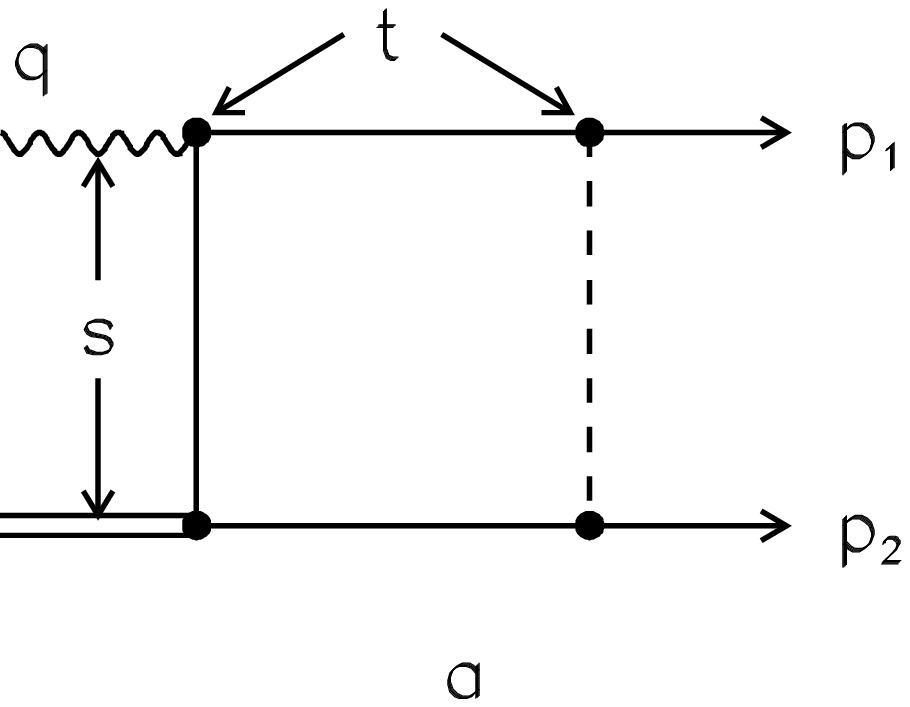,width=4cm}\hspace{1cm}
            \epsfig{file=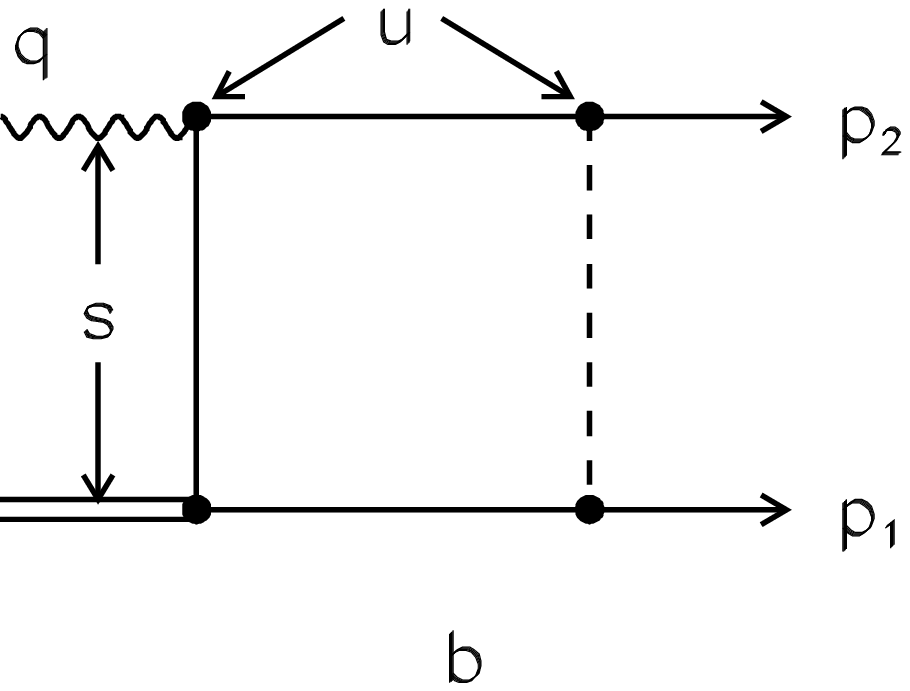,width=4cm}\hspace{1cm}
            \epsfig{file=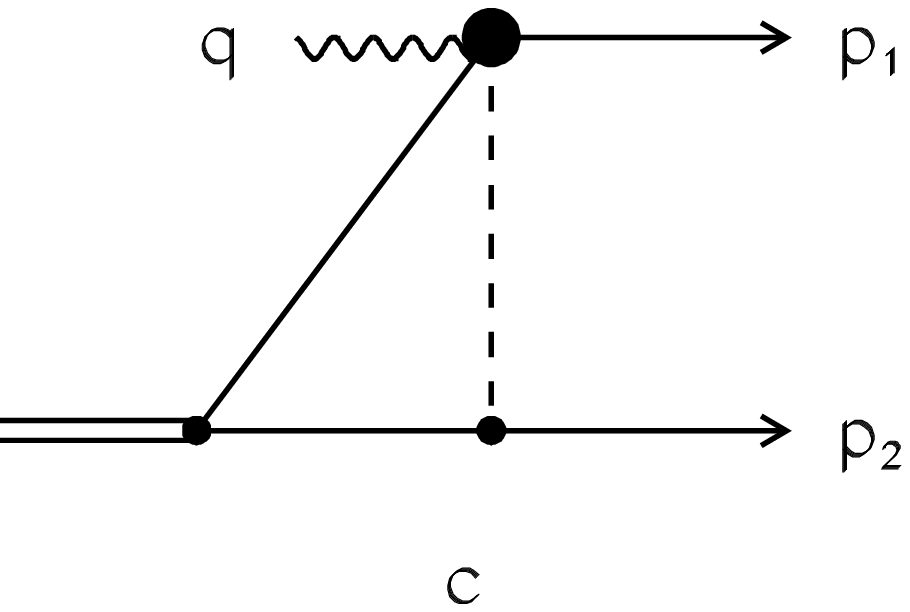,width=4cm}}
\vspace{1cm}
\centerline{\epsfig{file=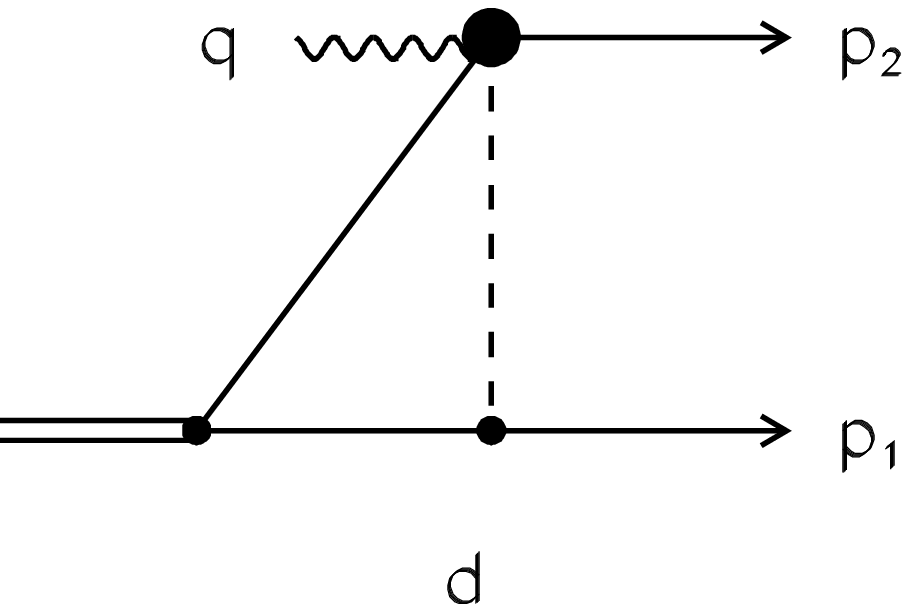,width=4cm}\hspace{1cm}
            \epsfig{file=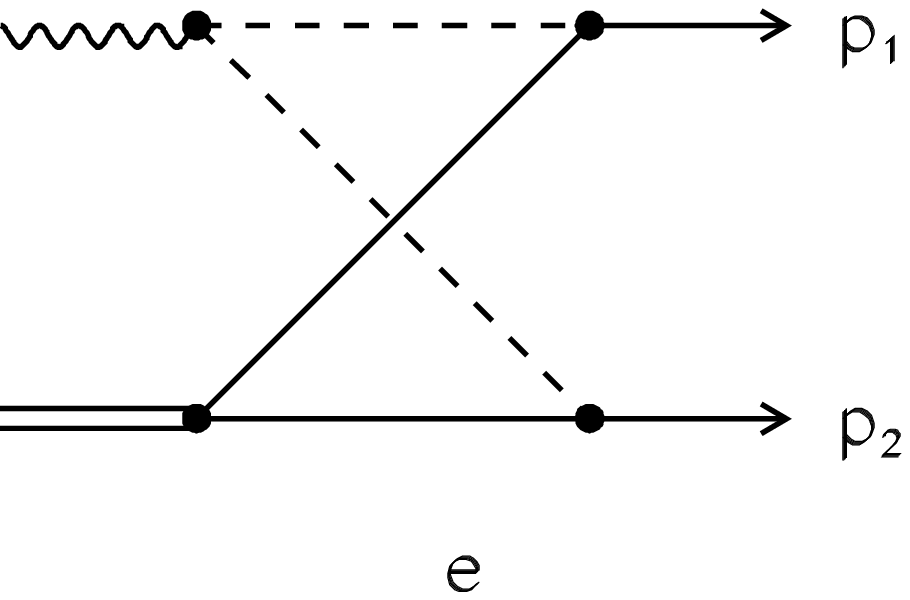,width=4cm}\hspace{1cm}
            \epsfig{file=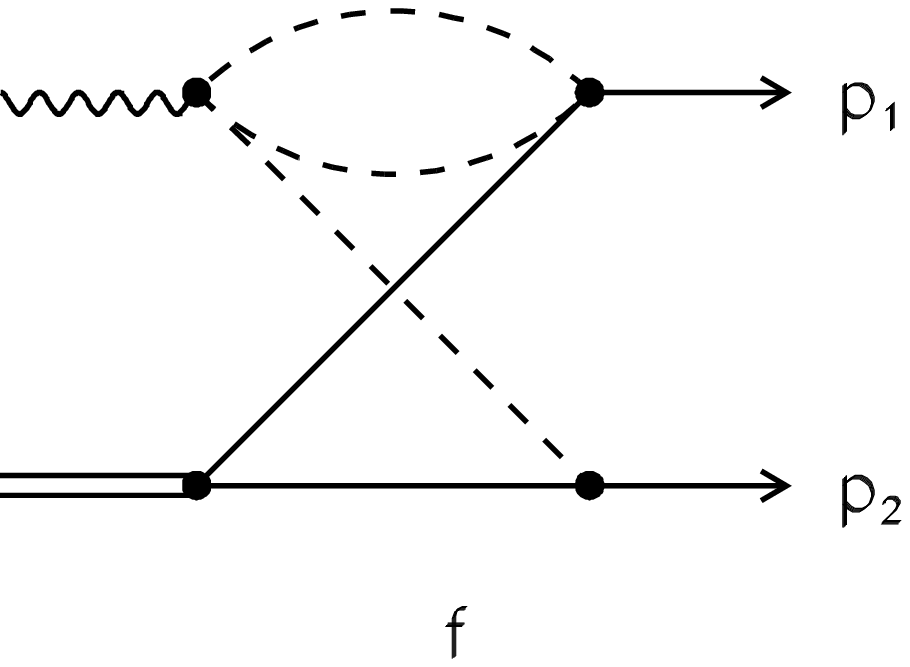,width=4cm}}
\vspace{1cm}
\centerline{\epsfig{file=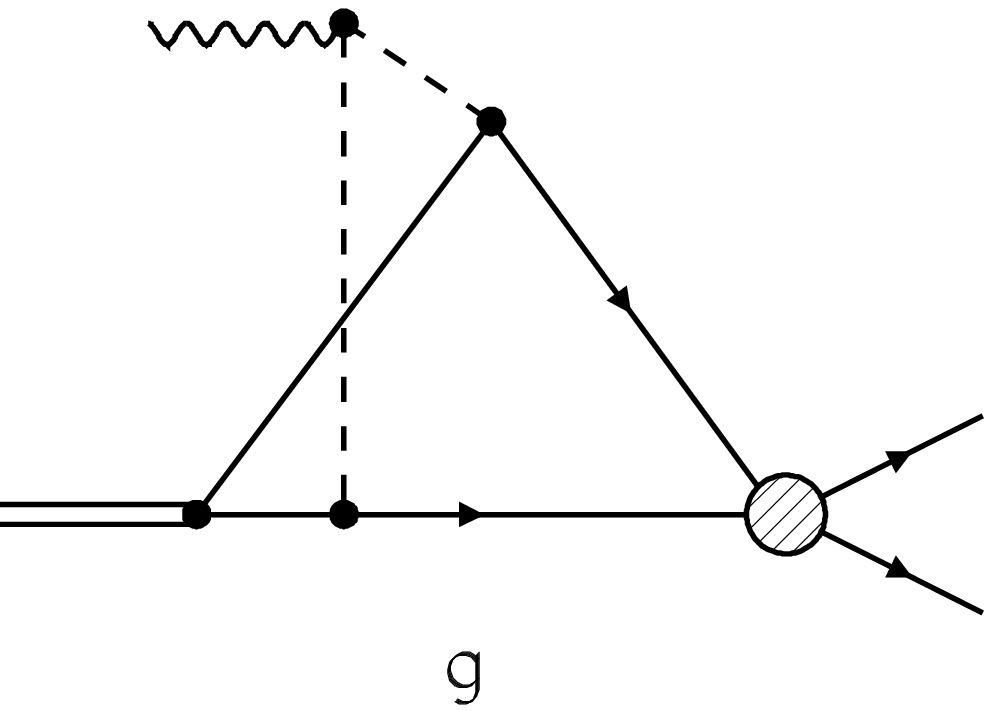,width=4cm}\hspace{1cm}
            \epsfig{file=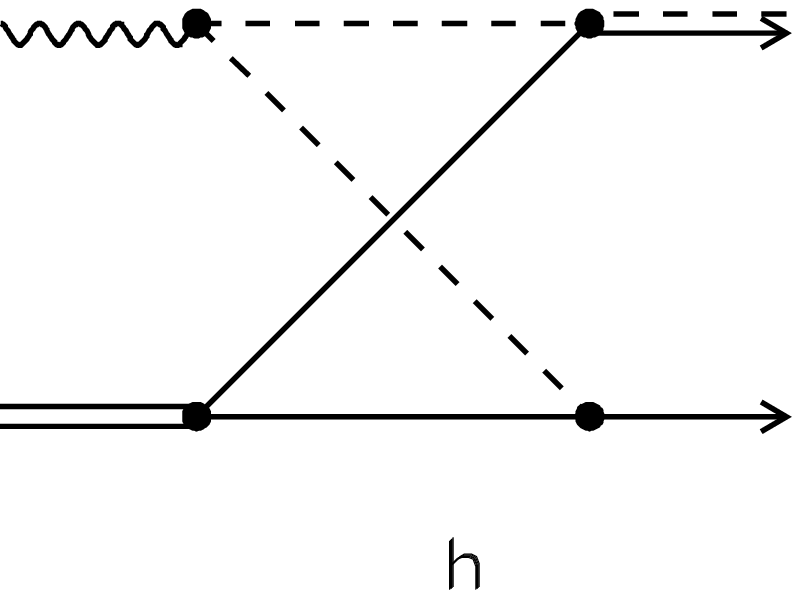,width=4cm}\hspace{1cm}
            \epsfig{file=f16g.eps,width=4cm}}
\caption{}
\end{figure}

\begin{figure}
\centerline{\epsfig{file=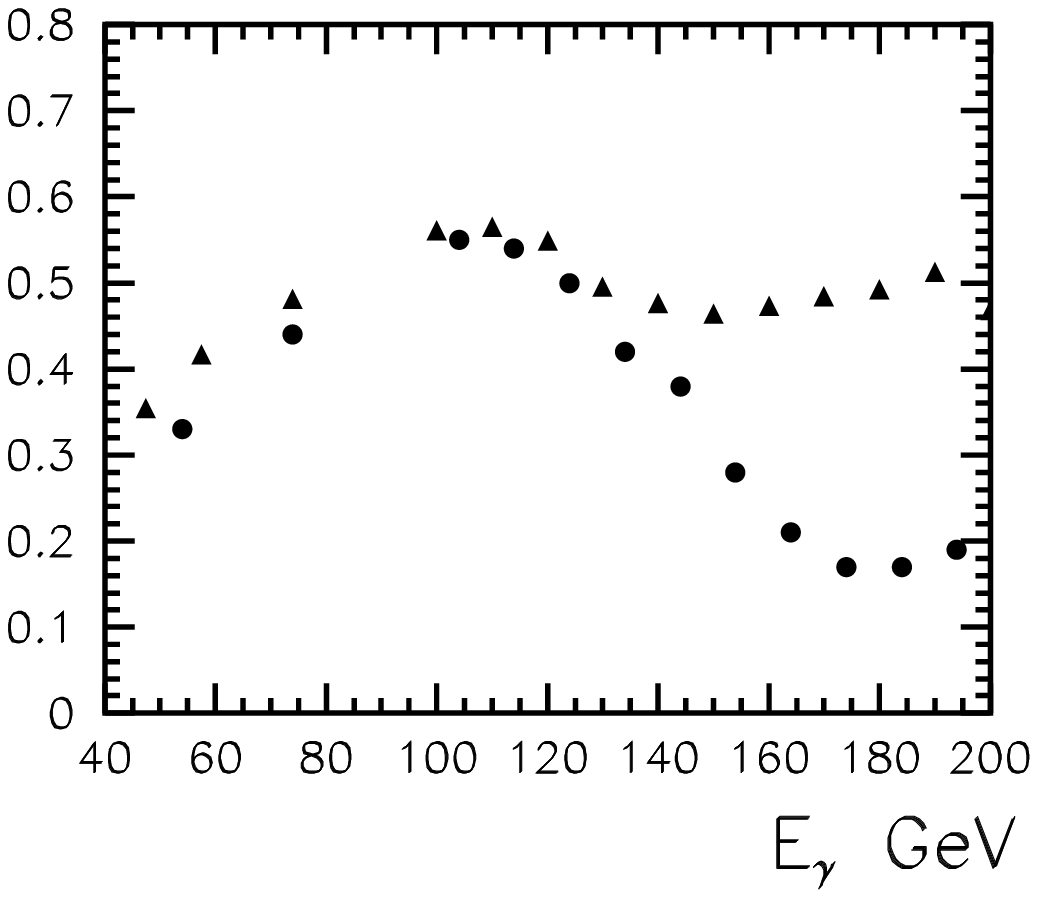,width=10cm}}
\caption{}
\end{figure}

\end{document}